\documentstyle[graphicx]{l-aa}
\onecolumn

\def\Eq#1{{Eq.~(\ref{e:#1})}}	
\def\Ep#1{{~(\ref{e:#1})}}	
\def\Det{{\rm Det}} 		
\def\Fig#1{{Fig.~(\ref{f:#1})}}	
\def\Figs#1{{Figs.~(\ref{f:#1})}}	
\def\Fip#1{{~(\ref{f:#1})}}	
\def\be{\begin{equation}}
\def\ee{\end{equation}}
\def\ba{\begin{eqnarray}}
\def\ea{\end{eqnarray}}
\def\para{\parallel}
\def\ort{\bot}
\def\vk{{\bf k}}
\def\vx{{\bf x}}
\def\vx{{\bf x}}
\def\vq{{\bf q}}
\def\vu{{\bf u}}
\def\d{{\rm d}}
\def\i{{\rm i}}
\def\mg{\big <}
\def\md{\big >}
\def\grad{\nabla}
\def\lm{\lambda_{\rm max}}
\def\vdl{{\bf dl}}
\def\Eqs#1{{Eqs.~(\ref{e:#1})}}	
\def\Tab#1{{Table~(\ref{t:#1})}}	
\def\Tap#1{{~(\ref{t:#1})}}	

\def\disp{\displaystyle}

\begin{document}

\thesaurus{2 (12.12.1; 12.04.1; 12.03.4; 11.06.1)} 

\title{Vorticity generation in large-scale structure caustics}

\author{ C.~Pichon \inst{1,2,3}   \and F.~Bernardeau\inst{3,4}}
\institute{ CITA, 60 St. George Street, Toronto, Ontario M5S 1A7,
Canada,
\and
  Astronomisches Institut  Universitaet Basel, Venusstrasse 7
 CH-4102 Binningen Switzerland,
\and
  Institut d'Astrophysique de Paris, 98 bis Boulevard d'Arago, 75014 Paris,
 \and Centre d'\'etude	 de	Saclay,	Service	de Physique
  Th\'eorique, 91191 Gif-sur-Yvette, France 
  }

\offprints{C. Pichon: \\
pichon@astro.unibas.ch}
\maketitle


\begin{abstract}

A fundamental hypothesis for the interpretation of the measured large-scale
line-of-sight peculiar velocities of galaxies is that the large-scale cosmic
flows are irrotational.  In order to assess the validity of this assumption,
we estimate, within the frame of the gravitational instability scenario, the
amount of vorticity generated after the first shell crossings in large-scale
caustics.  In the Zel'dovich  approximation the first emerging singularities
form sheet like structures.   Here we compute  the expectation profile  of an
initial overdensity   under the constraint  that  it goes  through its first
shell crossing at the present time.  We find that this profile corresponds to
rather oblate structures   in  Lagrangian space.   Assuming the   Zel'dovich
approximation  is   still adequate not   only at   the first stages   of the
evolution but also  slightly after the  first shell crossing, we calculate the
size and shape of  those caustics and  their vorticity content as a function
of time and for different cosmologies.

The average vorticity created in these caustics is small: of the order
of one (in units of the Hubble constant).  To illustrate this point we
compute the   contribution    of such caustics     to  the probability
distribution function of the  filtered vorticity at  large scales.  We
find that this contribution that this yields a negligible contribution
at the 10  to 15 $h^{-1}$Mpc  scales.  It becomes significant only  at
the scales of 3  to 4 $h^{-1}$Mpc,  that is, slightly above the galaxy
cluster scales.


\end{abstract}

\section{Introduction} \label{s:intro}

The analysis of large-scale cosmic flows  has become a  very active field in
cosmology (see Dekel  1994 for a recent  review  on the subject).   The main
reason  is  that   it can  in principle  give   access  to direct  dynamical
measurements of various quantities  of cosmological interest.  There are now
a very large   number  of methods and   results  for the comparison of   the
measured large--scale  flows   with the  measured  density  fluctuations  as
observed in the galaxy catalogues.  Most of these methods are sensitive to a
combination of the density of the universe in units of the critical density,
$\Omega$, and the  linear bias, $b$, associated to  the mass tracers adopted
to  estimate  the   density    fluctuations.    The estimated  values     of
$\beta=\Omega^{0.6}/b$ are about $0.3$  to 1 depending  on the method or  on
the tracers that are used.  There are other lines of  activities that aim to
estimate $\Omega$  from only the {\it intrinsic}  properties of the velocity
field,  (i.e., without   comparison    with  the observed   galaxy   density
fluctuations).  All these  methods exploit non-Gaussian features expected to
appear in the velocity field, either the maximum expansion rate of the voids
(Dekel \&  Rees 1994),  non-Gaussian general  features as expected  from the
Zel'dovich approximation  (Nusser \&  Dekel 1993), or   the skewness of  the
velocity divergence  distribution (Bernardeau et  al.  1995).  Yet  they all
also assume that the  velocity field is {\it potential}.   This is  indeed a
necessary requirement   for  building the  whole  3D   velocity out  of  the
line-of-sight  informations   in   reconstruction schemes   such   as Potent
(Bertschinger  \& al. 1990,  Dekel et al. 1994).   This is also a required
assumption for  carrying calculations  in    the framework of   perturbation
theory.  It is therefore interesting to check the  rotational content of the
cosmic  flows at scales at  which they are  considered in galaxy catalogues,
that is at  about   10 to 15$h^{-1}$Mpc.   This  investigation  ought  to be
carried in the frame of the gravitational instability scenario with Gaussian
initial  conditions.   It is  known  that in   the  single stream  r\'egime,
primordial vorticity  is diluted by the  expansion and that the higher order
terms in a perturbation expansion cannot create ``new'' vorticity.  Hence it
is natural to assume that the vorticity on  larger scales originate from the
(rare) regions where multi-streaming occurs.   During the formation of large
scale structures  this happens first   when the largest caustics  cross  the
first singularity,  creating  a  three-flow  region where vorticity   can be
generated.  As we argue in Sect.~2, analytical calculations of constrained
random Gaussian fields suggest that  the largest  caustics that are  created
are sheet-like structures,   in  rough  agreement  with what   is found   in
numerical simulations or in  galaxy catalogues.  It is therefore  reasonable
to use  Zel'dovich's approximation  to describe the subsequent evolution of
those objects.

In order  to estimate the large scales  vorticity  distribution we therefore
proceed in five  steps: first we evaluate  the mean constrained random field
corresponding to  a local  asymmetry of the  deformation  tensor on  a given
scale, $R_{L}$; secondly  we solve for   the multi-flow r\'egime  within the
generated caustic,  using   Zel'dovich's  approximation  throughout,  even
slightly  beyond this  first singularity.   We  then evaluate the  vorticity
field in that caustic. The next step involves modelling the variation of the
characteristics of typical caustics as a function of time for different power
spectra. Finally, we  estimate the amount  of vorticity expected at large
scales arising from   large scale flow caustics.

For the sake of simplicity and because is  pedagologically more appealing,  we
present calculations carried out  in two  dimensions as   well   as in   three
dimensions. The former case is  in particular easier to handle numerically.

The second Section of  this paper evaluates  the characteristics of the typical
caustics expected at  large--scale in a 2D  or 3D density field.   The third
Section is devoted  to the explicit calculation of  the vorticity for the  most
typical caustics.  The fourth Section provides an estimate for the shape of the
tail of  the   probability  distribution  function of  the   modulus  of the
vorticity in a sphere of a given radius. It is followed by a
discussion on the validity and implications of these results.

\section{Asymmetric constrained random fields} \label{s:random}

Since it is not our ambition to solve the problem  of deriving the vorticity
statistics  in its  whole generality  the  vorticity will be estimated  only
within specific  but  typical caustics in  the  framework of the  Zel'dovich
approximation. 

 The  first  step involves building   an initial density  field   in which a
caustic will eventually appear.  The initial fluctuations  are assumed to be
Gaussian  with a given  power spectrum $P(k)$,  characterizing the amplitude
and shape of the initial fluctuations.  No  {\it a priori} assumptions about
the  values of $\Omega$ and  $\Lambda$ are made.   It will be shown that the
statistics has very straightforward  dependences upon these parameters.  The
expectation values of the  random variables, $\delta(\vk)$, corresponding to
the Fourier transforms of the local density field,
\begin{equation}
\delta(\vx)=\int \d^3\vk\,\delta(\vk)\,\exp[\i \vk\cdot\vx],
\label{e:aaa} \end{equation}
are  calculated once a  local constraint has  been imposed.  This constraint
will be chosen  so that the caustic-to-be  will have just gone through first
shell crossing at  the present time. It is  expressed in terms of the  {\it
local} deformation   matrix   of the {\it   smoothed}   density  field.  The
components of the local deformation tensor at the position $\vx_0$ are given
by
\begin{equation}
\Phi_{i,j}(\vx_0)=\int\d^3\vk\      \delta(\vk)\      W_D(k\,   R_L) \exp[\i
\vk\cdot\vx_0]\, {\vk_i\vk_j\over k^2},
\label{e:aab} \end{equation}
where $W_D$ is the adopted window function. In what follows,
we will use the top-hat window function for which
\be
W_2(k)=2\ {J_1(k) \over k^{1/2}}\quad {\rm in\ }\ \ 2D,\ \ 
W_3(k)=3\sqrt{\pi/2}
\ {J_{3/2}(k) \over k^{3/2}}\ \ {\rm in\ }\ \ 3D,\label{e:w3} 
\end{equation}
where $J_{\nu}$ is the Bessel function of index $\nu$.
The scale $R_L$ is the scale of the caustic in Lagrangian
space. Here $\sigma_0$  stands for the {\it rms} density fluctuation
at this scale:
\begin{equation}
\sigma_0^2 =\int\d^3k\,P(k)\,W_2^2(k\,R_L).
\label{e:aac} \end{equation}
For the sake of simplicity  a typical caustic is  chosen to be characterized
by the average local  perturbation over a sphere of  radius $R_L$ for  which
the deformation tensor at its centre is fixed.  We are aware that this is
a  somewhat drastic  approximation but  consider that,  at large scales, the
behaviour of caustics having the mean initial profile will be representative
of the average behaviour.  This is certainly  not true at small scales where
the complex interactions of structures at different scales and positions are
expected to affect the global behaviour of any  given caustic. For some rare
enough objects however we expect the fluctuations around the mean profile to
be small enough to affect only weakly the global properties of the caustics.
This  has been shown  to be  true in the   early stages of the  dynamics for
spherically symmetric perturbations (Bernardeau 1994a).  In the following we
will, however, encounter  properties (see  \S   3.3) that we think  are  not
robust against small scale fluctuations.  Such properties will be ignored in
the subsequent applications of our results.

Within  the frame of  this calculation,  the  values of $\delta(\vk)$  hence
correspond to the expectation values of $\delta(\vk)$ for the power spectrum
$P(k)$ when  the constraints on the deformation  tensor are satisfied. These
solutions can be   written as a  linear  combination  of the values  of  the
deformation tensor:
\begin{equation}
\delta(\vk)= 
\sum_{i=1}^D-{\left(C^{-1}\right)_{0,i}\over\left(C^{-1}\right)_{0,0}}
\ \lambda_i \, 
\equiv \sum_{i=1}^D\,\,\alpha_i\,\lambda_{i}\, ,
\label{e:deltavk} \end{equation}
where the coefficients $C$ is  the matrix
of the cross-correlations  between the random Gaussian variables $\Phi_{ij}$
and $\delta(\vk)$ as shown in Appendix~\ref{s:avg}.  
 In   \Eq{deltavk} the summation  is  made  only on  the
diagonal  elements of the deformation tensor  since it is always possible to
choose the axis   in such a  way that   the  other elements  are zero.
In this instance, the diagonal elements are identified with the
eigenvalues  $\lambda_i$, of the matrix.

\subsection{The 2D field}

In 2D geometry, the two  coefficients  $\alpha_1$ and $\alpha_2$ defined  by
\Eq{deltavk} are given by
\begin{eqnarray}
\alpha_1=(3 I_1-I_2)/\sigma_0^2\, ,\ \ 
\alpha_2=(3 I_2-I_1)/\sigma_0^2\, ,
\label{e:aad} 
\ \ {\rm where } \ \ 
I_i=\mg\delta_k\,\Phi_{ii}\md=P(k)\,
W_2(k R_L)\,{\vk_i^2\over k^2}.
\label{e:aae} \end{eqnarray}
The brackets, $\mg.\md$, denote ensemble
averages over the initial (unconstrained) random density field.
As a result, \Eq{deltavk} reads 
\begin{equation}
\delta(\vk)=
{P(k)\,W_2(k R_L)\over\sigma_0^2}\,
\big[2\,(\lambda_1+\lambda_2)+4\,(\lambda_1-\lambda_2)\,
\cos(2 \theta)\big] \, ;
\label{e:deltak2D} 
\end{equation}
$\lambda_1$ and $\lambda_2$  are the eigenvalues  of  the deformation tensor
and where $\theta$ is the angle between $\vk$ and the eigenvector associated
with the  first eigenvalue (see  Appendix~\ref{s:avg} for details). Consider
the parameter $a$ defined by
\begin{equation}
a={2(\lambda_1-\lambda_2)\over\lambda_1+\lambda_2}.
\label{e:afe} 
\end{equation}
The  coefficient $a$ represents the  amount of  asymmetry in the fluctuation
(thus  $a=0$  corresponds to a spherically   symmetric  perturbation).  This
parameter is similar to  the eccentricity, $e$,  that was used by Bardeen et
al.  (1986)   and more specifically  by  Bond \& Efstathiou (1987)   for 2D
fields. In  these studies however   investigations were  made for
the  shape of the peaks  around the maximum (i.e.  eigenvalues of the second
order   derivatives  of the   local density),  so    $a$ and  $e$ cannot  be
straightforwardly identified.

\begin{figure}
\begin{minipage}[b]{0.45 \linewidth}
\centering \includegraphics[width=3.in]{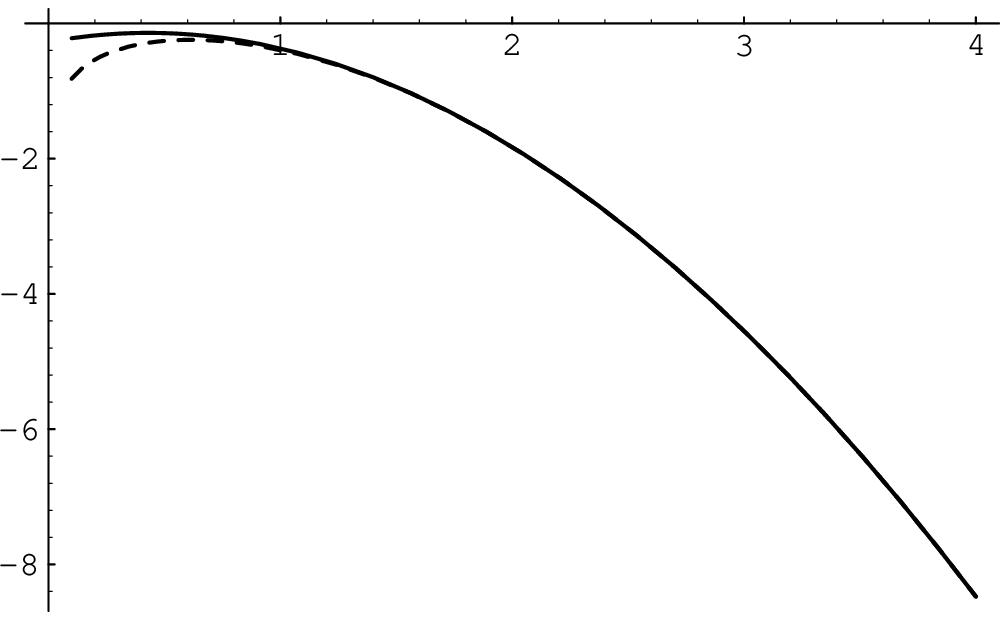}
\caption{The  distribution function of  $\lm/\sigma_0$   (solid line) in  2D
dynamics. The dashed line is given by (\Eq{p2Dlm}):
}
\label{f:fig-1}
\end{minipage}
\begin{minipage}[b]{0.2 \linewidth}
\end{minipage}
\begin{minipage}[b]{0.45 \linewidth}
\centering \includegraphics[width=3.in]{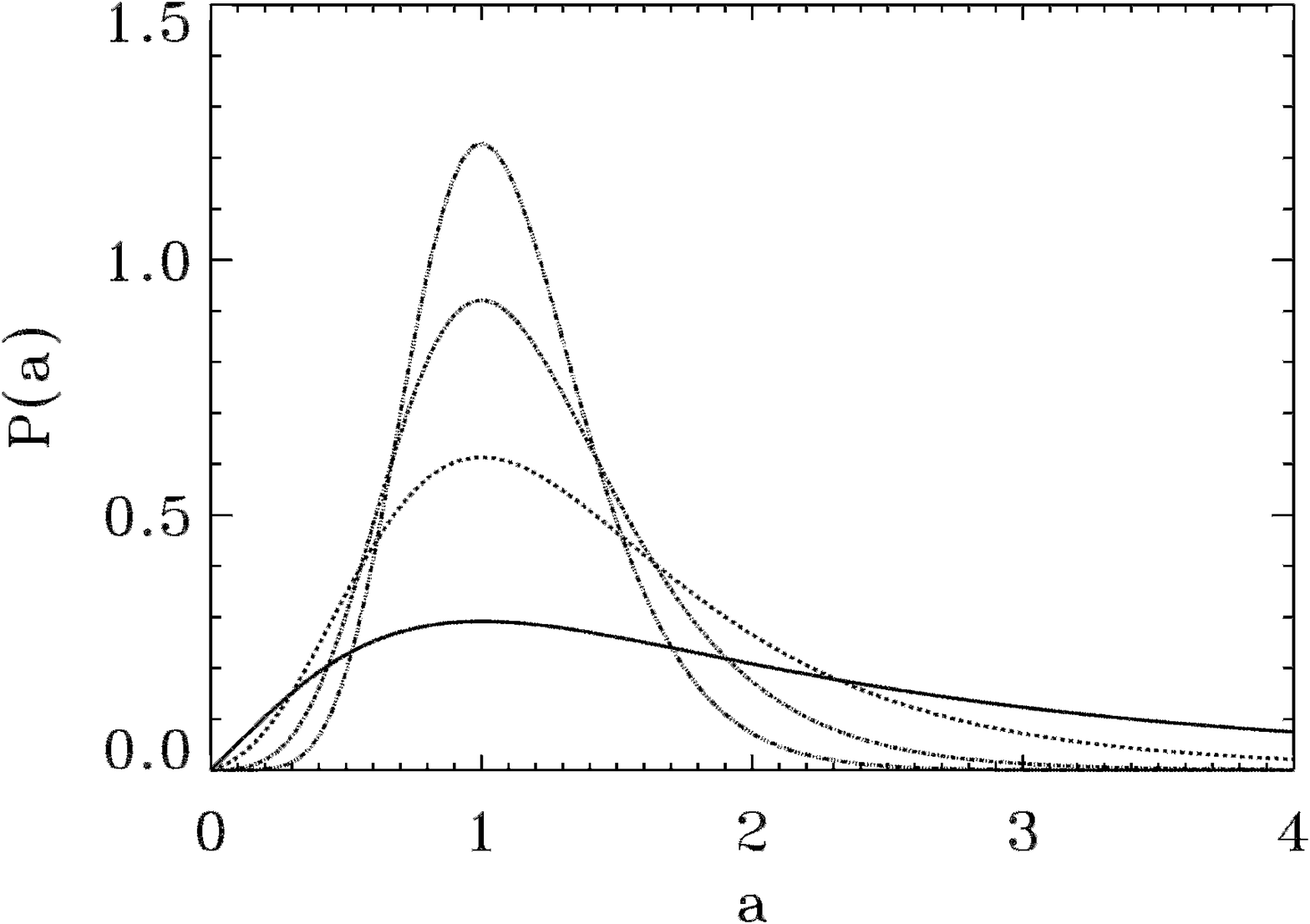}
\caption{The  distribution  functions    of  $a$  for   fixed   values    of
$\lm/\sigma_0=1,\ 2,\   3,\ 4$ (respectively  the solid,  long dashed, short
dashed and long dotted dashed lines).
}
\label{f:fig-2}
\end{minipage}
\end{figure}
The  formation time  of the   first  singularity is
determined by the maximum value of  the eigenvalues, $\lm$.  It is therefore
relevant   to calculate   the  distribution   function  of   $\lm$, and  the
distribution  function of  $a$  once $\lm$ is  known.   From the statistical
properties of the matrix elements  $\Phi_{ij}$ we derive the
distribution function of the  eigenvalues $\lambda_{\rm min}$ and $\lm$ (see
Appendix~\ref{s:df}), which reads
\begin{eqnarray}
P(\lambda_{\rm min},\lm)&=&
{2^{3/2}\over\pi^{1/2}\,\sigma_0^3}
\,(\lm-\lambda_{\rm min}) 
\exp\left[-{1\over\sigma_0^2}\left({3\over2} 
J_{1}^2-{4}\,J_{2}\right) \right],
\label{e:p2D} \end{eqnarray}
with
\begin{eqnarray}
J_{1}=\lambda_{\rm min}+\lm \, , \quad  
J_{2}=\lambda_{\rm min}\,\lm.
\label{e:fit2D} \end{eqnarray}
The distribution function  of  $\lm$ follows  by  numerical integration over
 $\lambda_{\rm min}$.  \Fig{fig-1} shows the  distribution function of $\lm$
 in units of  the variance.   The  dashed line corresponds  to the approximation, valid at
$\lambda_{\rm max} / \sigma_0 \gg 1$:
\begin{eqnarray}
p_{\rm max}
(\lm)\,\d\lm
\approx 1.5\,{\lambda_{\rm  max}\over\sigma_0}
\exp\left[-{4\over3} \left({\lm\over\sigma_0}\right)^2\right]\ 
{\d\lm\over\sigma_0}.
\label{e:p2Dlm} \end{eqnarray}
The distribution function of $a$  for different values of $\lm/\sigma_0$  is
presented in  \Fig{fig-2}.  It  turns out  that the most  significant  value
corresponds to $a\approx 1$. In  the following this value  is chosen as  the
typical value for the asymmetry in two dimensions.

\subsection{The 3D field} \label{s:3Dfield}

In three dimensions the 
geometry is slightly more complicated and yields for the constrained density
field
(see Appendix~\ref{s:df} for details)
\begin{eqnarray}
\delta(\vk)&=&
{3\ P(k)\,W_3(k\,R_L)\over 8\sigma_0^2}\ 
\Big(
\lambda_1\,\big[1+5\,\cos(2\phi_k)-5 \cos(2\theta_k)-
5\,\cos(2\phi_k)\,\cos(2 \theta_k)\big]+  \nonumber\\&&
\lambda_2\,\big[1+5\,\cos(2\phi_k)-5 \cos(2 \theta_k)-
5\,\cos(2\phi_k)\,\cos(2 \theta_k)\big]
+\,2\lambda_3\,\big[3+5\,\cos(2\,\theta_k)\big]
\Big),
\label{e:deltak3D} \end{eqnarray}
where $\theta_k$ and  $\phi_k$  are polar angles  of  the vector $\vk$  with
respect  to the    basis of  the    eigenvectors  associated to   the  three
eigenvalues,    $\lambda_1,\lambda_2,\lambda_3$.   The asymmetry    of   the
distribution is again characterized by the values of
\begin{equation}
a=5\,
{2\lambda_3-\lambda_1-\lambda_2\over \lambda_1+\lambda_2+6
\lambda_3}\, , 
\ \ {\rm and} \ \
b=5\,{\lambda_1-\lambda_2\over \lambda_1+\lambda_2+6\lambda_3}.
\label{e:aaj} 
\end{equation}
When $b$ only  is  zero \Eq{aaj} corresponds  to  a perturbation with  axial
symmetry, and when both $a$ and  $b$ are zero  it is a spherically symmetric
perturbation.  In terms of $a$ and $b$ \Eq{deltak3D} then becomes
\begin{eqnarray}
\delta(\vk)=
{3\,P(k)\,W_3(k\,R_L)\over 8\,\sigma_0^2}\,(\lambda_1+\lambda_2+6\lambda_3)
\Big(
1+a\,\cos(2 \theta_k)+b\,\cos(2\phi_k)\big[1+\cos(2 \theta_k)\big]\Big).
\label{e:aak} \end{eqnarray}
Let us now  evaluate the distribution of $a$  and $b$ from  the distribution
function  of   the   eigenvalues  $(\lambda_1,\lambda_2,\lambda_3)$  in   3D
(assuming $\lambda_1>\lambda_2>\lambda_3$) in order to identify the shape of
the most significant caustics.  This  distribution is given by (Doroshkevich
1970)
\begin{eqnarray}
P(\lambda_1,\lambda_2,\lambda_3)=
{5^{5/2}\ 27\over8\,\pi\,\sigma_0^6}\,(\lambda_1-\lambda_2)\,
(\lambda_1-\lambda_3)\,(\lambda_2-\lambda_3)\,
\exp\left[-{1\over\sigma_0^2}\left(3 J_{1}^2-{15\over2}\,J_{2}\right) 
\right],
\label{e:p3D} 
\end{eqnarray}
with
\begin{eqnarray}
J_{1}=\lambda_1+\lambda_2+\lambda_3 \, , \ \ {\rm and } \ \
J_{2}=\lambda_1\lambda_2+\lambda_2\lambda_3+\lambda_3\lambda_1.
\label{e:aam} \end{eqnarray}

\begin{figure}
\begin{minipage}[b]{0.45 \linewidth}
\centering \includegraphics[width=3.in]{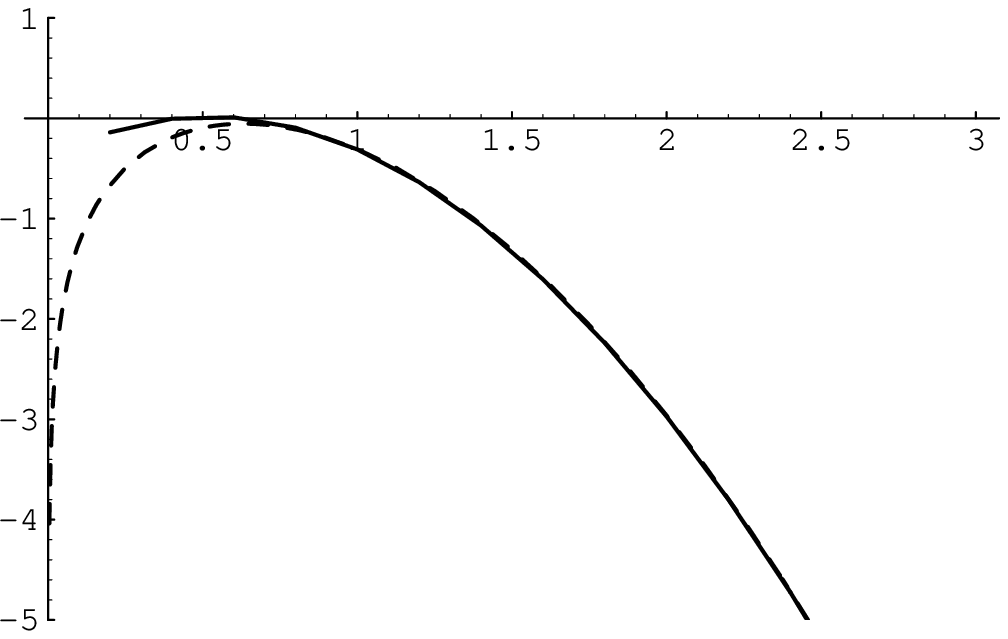}
\caption{The distribution function of $\lm/\sigma_0$ 
(solid line)
in 3D dynamics. The dashed line is the analytical fit (\ref{e:p3Dlm}).
}
\label{f:fig-3}
\end{minipage}
\begin{minipage}[b]{0.3 \linewidth}
\end{minipage}
\begin{minipage}[b]{0.45 \linewidth}
\centering \includegraphics[width=3.in]{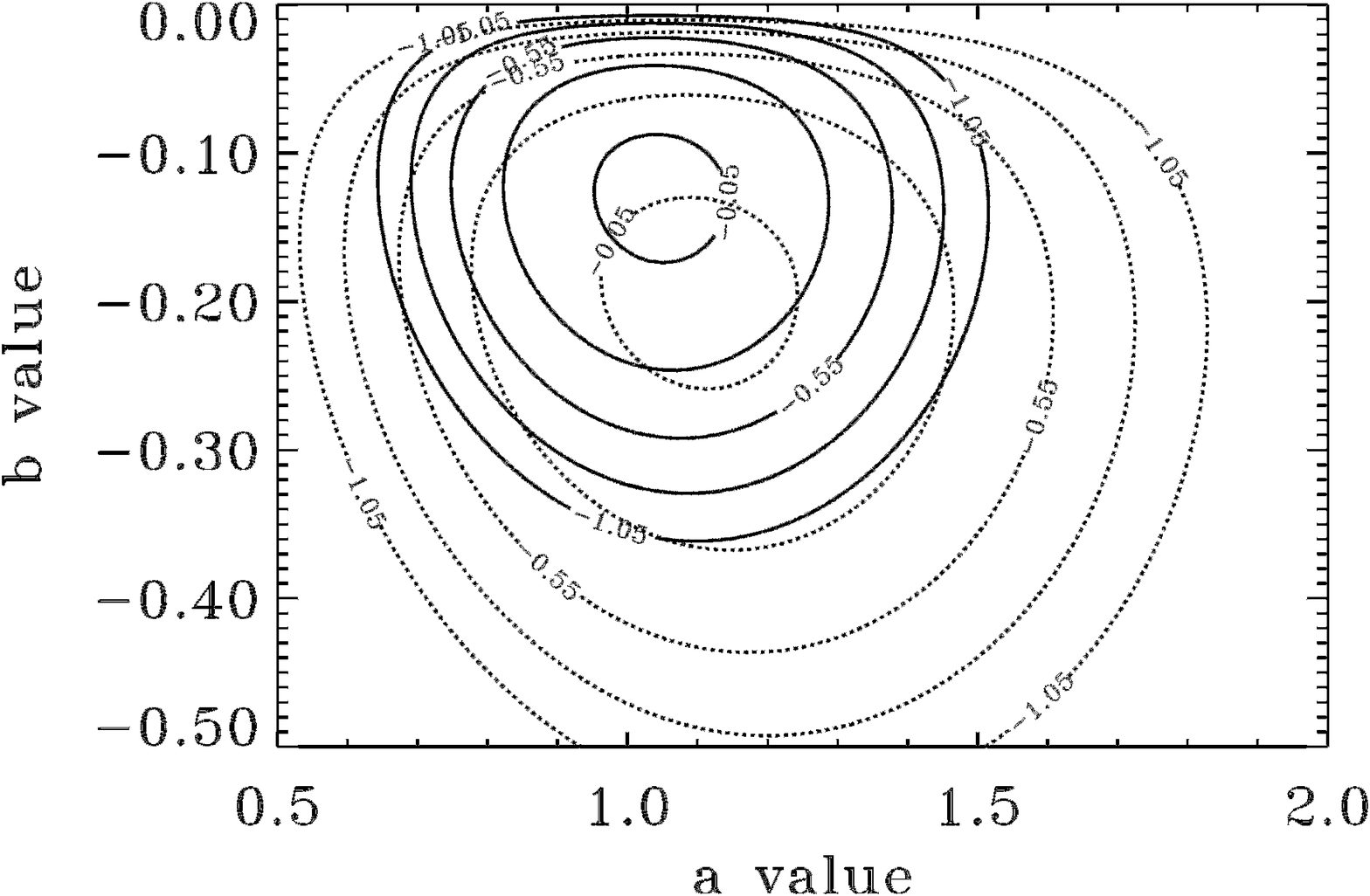}
\caption{The contour plot for the distribution of $a$ and $b$
for a fixed value of $\lm/\sigma_0=2$ (dashed lines) and
$\lm/\sigma_0=3$ (solid lines). The lines are evenly distributed in 
a logarithmic scale.
}
\label{f:fig-4}
\end{minipage}
\end{figure}

From this expression we compute numerically the distribution function of the
maximum  eigenvalue (\Fig{fig-3}).  An analytical  fit  of this distribution
function is provided by its   behaviour at large $\lm$
\begin{eqnarray}
p_{\rm max}(\lm)\,\d\lm
\approx 6\,\left(\lambda_{\rm  max}\over\sigma_0\right)^2\,
\exp\left[-{5\over2} \left({\lm\over\sigma_0}\right)^2\right]
\,{\d\lm\over\sigma_0}.
\label{e:p3Dlm} \end{eqnarray}
This fit is accurate for the rare event tail (as shown in \Fig{fig-3}), which
will be relevant  for the derivation  of Sect.~\ref{s:stat}.  For  a given
value of $\lm$  we compute the distribution of   the other eigenvalues,  and
thus the join distribution function of $a$ and $b$.

The   resulting  contour   plot    corresponding  to   $\lm/\sigma_0=2$  and
$\lm/\sigma_0=3$ is illustrated on \Fig{fig-4}.   As for the distribution of
$a$ in the previous subsection in 2D it depends only weakly upon the adopted
value of $\lm$  (although the position of the  maximum varies a little), and
it tends to be all the more peaked  on its maximum  as $\lm$ is large.  This
implies that a typical caustic  will be given  by $a\approx 1$ with a  small
$b$.  For further simplifications we will assume  that $b=0$. Such a caustic
then corresponds to a pancake-like structure with  axial symmetry. Note that
this  result seems to  differ from the results of  Bardeen et al. (1986) who
found  that  the shape of the   rare  peaks should   be somewhat spherically
symmetric or filamentary (this picture was recently sustained by Pogosyan \&
al., 1996,  from the   result   of $N$-body  simulations).   This   apparent
discrepancy is due  to the constraint under which  the expectation values of
$a$ and  $b$ are calculated.   In Bardeen  et al.'s  work the constraint  is
given  by the value    of the local density,  i.e.    the sum of  the  three
eigenvalues,  whereas  in  this paper  we put  a  constraint  on the largest
eigenvalue.  This  is a natural  assumption for this investigation since the
multi-streaming  occurs as soon  as  a singularity has  been  reached in one
direction.      Of  course,  this  analysis   assumes    that the Zel'dovich
approximation holds    in order to   predict  the time  at which  this first
singularity is  reached.   For oblate initial structures  such   as the ones
obtained for the most  likely values of $a$ (see  Figs.  5 and 6), we expect
that this approximation is sufficiently accurate.

\section{The  geometry  \& vorticity of large-scale caustics}
\label{s:geom}

In this  section  we investigate the  properties of   the caustics  that are
induced by the initial density fluctuation profiles we found in the previous
section. All the  calculations are  performed within  the framework  of  the
Zel'dovich approximation, even sightly after the first shell crossing.

\subsection{The linear displacement field}

In the framework of the Zel'dovich approximation
the displacement field can be written
\begin{equation}
\vx=\vq+D(t)/D(t_0)\,\Psi(\vq) \, ;
\label{e:displacement} \end{equation}
where $D(t)$ accounts for the time dependency of the linear growing mode (it
is  proportional to the  expansion factor  in case  of an Einstein-de Sitter
geometry only).  An  important simplification is that, at  the order  of the
Zel'dovich approximation, this displacement field is  separable in time and
space, and its space dependence, $\Psi(\vq)$, is potential, i.e., there is a
velocity potential $U(\vq)$ so that
\begin{equation}
\Psi(\vq)=\grad_{\vq}\cdot U(\vq)\, .
\label{e:aao} \end{equation}
This velocity potential is  given by
\begin{equation}
U(\vq)=\int\d^3\vq\,\delta(\vk)\,{1\over k^2}\,
\exp[\i \vk\cdot(\vq-\vq_0)] \, . 
\label{e:uq} \end{equation}
By  construction the point $\vq_0$  in  Lagrangian space  corresponds to the
point $\vx_0$ in real space (central position of the caustic).  Both of them
will be taken to  be zero.  For  the calculation of the explicit expressions
of $\delta(\vk)$ and $U(\vq)$ we will assume that the power spectrum follows
a power law behaviour,
\begin{equation}
P(k)\propto k^n,\label{e:index}
 \end{equation}
characterized by the power index $n$. 
From \Eq{index} the expression of the linear variance
as a function of scale follows
\begin{equation}
\sigma(R_L)\propto R_L^{-(n+D)/2}.
\label{e:sigRL} \end{equation}
This  approximation  is  valid within  a  limited  scale  range  as  will be
 discussed in Sect.~5. At the scales of interest the index $n$ is expected
 to  be  the range of   $n\approx-1,\ -2$ from  the  constraints
 obtained with
the large-scale galaxy catalogues, like the APM survey
(Peacock 1991) the IRAS galaxy survey (Fisher et al. 1993) or from
X-ray cluster number  counts (Henry \& Arnaud 1991, Eke et al. 1996, 
Oukbir \& Blanchard 1997).
In two dimensions  there are of course  no such
 observationally motivated  values, but we will  consider $n$ of the order of
 $-1$ as an illustrative case.

\subsubsection{The 2D potential}
From the \Eqs{deltak2D},\Ep{uq} it is possible to calculate the expression of 
the potential 
\begin{equation}
U(\vq)=G(0,n-2,q)+ 
 a\,\cos(2\theta_q)\,\left[G(0,n-2,q)-
2\ G(1,n-2,q)\right] \, , 
\quad {\rm with} \quad
G(\nu,n,q)=
\int\d^2\vk\,k^n\,{J_{\nu}(k\,q)\over (k\,q)^{\nu}}\,W_{2D}(k).
\label{e:2Du} \end{equation}
The latter expression is given by
\begin{equation}
G(\nu,n,q)=
\,_2F_1(1+n/2,n/2,1+\nu,q^2)  \,, \quad  {\rm for} \quad q<1 \,,
 \quad {\rm and}
\label{e:aaq} \end{equation}
\begin{eqnarray}
G(\nu,n,q)=
{\Gamma(1+\nu)\,\Gamma(1-n/2)\over q^{n+2}\,\Gamma(\nu-n/2)}\
\,_2F_1(1+n/2,1-\nu+n/2,2,q^{-2}) \, , \ \ {\rm for} \ \ q>1.
\label{e:aar} \end{eqnarray}
The expressions for the gradients  of the potential involve similar 
hyper-geometric functions.

\subsubsection{The 3D potential}
The expression of the potential  following from \Eqs{deltak3D},\Ep{uq}
becomes quite complicated,
but involves here only ``simple'' functions. It reads
\begin{equation}
U(\vq)=[V(q)-V(-q)]/q^3,
\label{e:3Du} \end{equation}
with
\begin{eqnarray}
V(q)&=& 
   \vert 1 + q \vert^{2-n}\,{\rm sign}(1 + q) \Big(
{{ {\it A}(q) - {\it B}(q)\,
        \big[ b\,\cos (2\,\phi )\,\left[ 1 - \cos (2\,\theta ) \right]  + 
          a\,\cos (2\,\theta ) \big] }} \Big) \, , \\
A(q)&=&-10\,{q^2} + 7\,n\,{q^2} - {n^2}\,{q^2} + 5\,{q^3} - n\,{q^3} 
+
  a\,\left( -1 + 2\,q - n\,q + 2\,{q^2} - n\,{q^2} - {q^3} \right) 
  \, , 
\\
B(q)&=&
3 - 6\,q + 3\,n\,q + 4\,{q^2} - 4\,n\,{q^2} + {n^2}\,{q^2} - 2\,{q^3} + 
  n\,{q^3}
\label{e:aar2} \end{eqnarray}
Note that the potentials in \Eqs{2Du} and \Ep{3Du}  have discontinuous
derivative at $q=1$,
which is an artifact of using a top-hat  window function.  Note also
that the potentials given here have arbitrary normalizations. This is 
of no   consequence  for the   derived results since the  global
normalization of the initial density profile
is absorbed   in  the discussion   for the  value of  $\lm$
(Sect.~\ref{s:stat}).

\subsection{The shape of the caustics} \label{s:shape}

\begin{figure*}
\begin{minipage}[b]{0.45 \textwidth}
\centering \includegraphics[width=3.in]{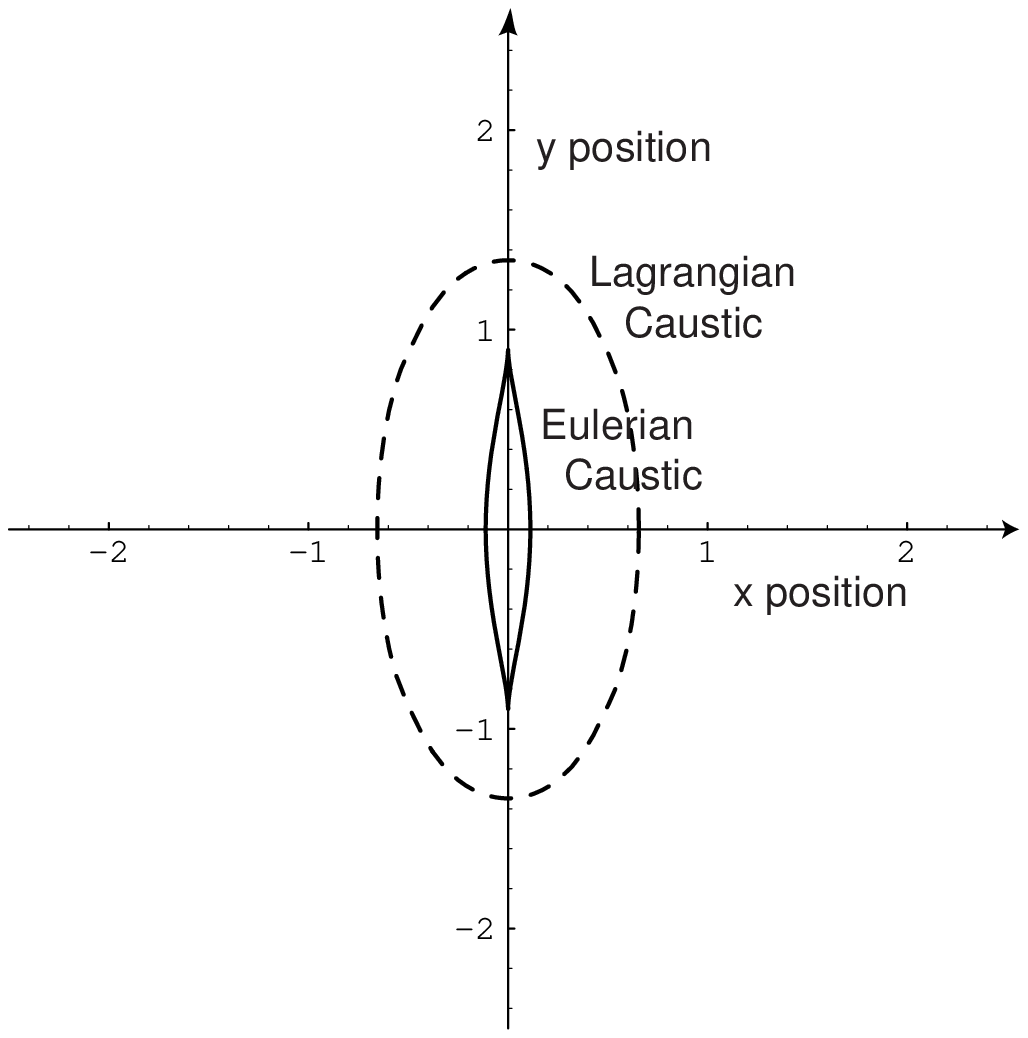}
\caption{
The shape of the caustic for the 2D dynamics, $n=-1$, and $\lm\approx 1.3$.
The dashed line is the shape in Lagrangian space and the
solid line the shape in real space.
}
\label{f:fig-5}
\end{minipage}
\begin{minipage}[b]{0.1 \linewidth}
\end{minipage}
\begin{minipage}[b]{0.45 \textwidth}
\centering 
\includegraphics[width=3.in]{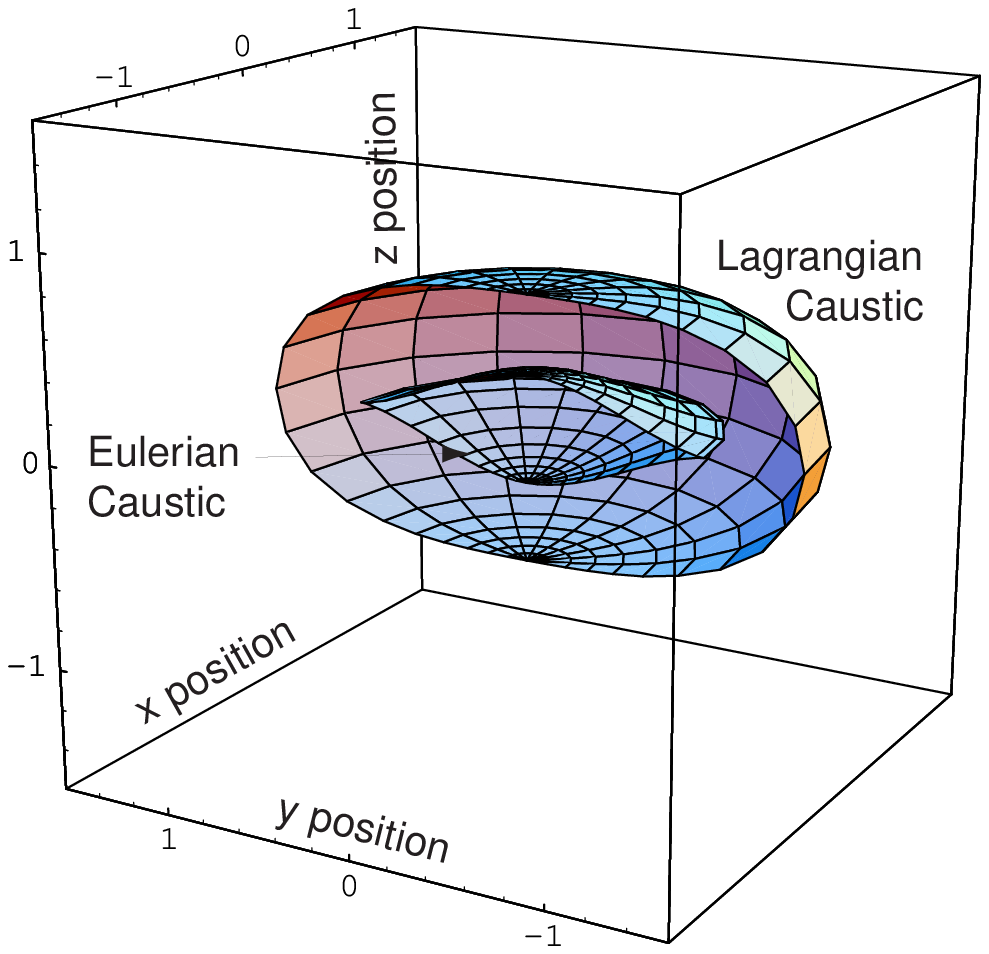}
\caption{The shape of the caustic for the 3D dynamics, $n=-1.5$
and $\lm\approx 1.5$.
The external shell is the Lagrangian position of the 
caustic, the internal one its position in real space.
}
\label{f:fig-6}
\end{minipage}
\end{figure*}

A  multi-flow region forms  as  soon as  \Eq{displacement} has more  than one
solution.   The corresponding region forms  the so-called caustic.   These regions are
illustrated in \Figs{fig-5}  and    \Fip{fig-6} in  respectively 2    and  3
dimensions for  typical values of the parameters.   The solid lines show in 2D the
shape of the caustic in real space, and the dashed lines  their shape in the
original Lagrangian space.

For the chosen values of $a$ and $b$ and for the relevant $\lm$ the caustics
form elongated structures.    These figures  are  plotted  in units   of the
smoothing  scale  $R_L$.   They suggest that  the  largest  dimension of the
caustics are roughly of  the  order of magnitude  of the  initial Lagrangian
scale.  Note that the boundaries of the  caustics correspond to surfaces (or
lines in 2D) where the Jacobian of the transformation between  Lagrangian
space and  real space vanishes, i.e.
\begin{equation}
J(\vq)=\big\vert{\partial\vx\over\partial\vq}\big\vert=0.
\label{e:aas} 
\end{equation}
The  size  and shape  of  these caustics   are characterized, in  2D  and 3D
(although only   approximately),  by two  lengths,  the  half-depth   of the
caustic, $d$, (that is the distance that has been covered by the shock front
after the first  singularity) and its half-extension  $e$.  For instance  in
\Fig{fig-5}  the value of $d$ is  about $0.1$ and  the value of $e$ is about
$0.9$ in units of the Lagrangian size of  the fluctuation $R_L$. In the case
of the 3D dynamics $e$  corresponds to the  radius of  the caustic since  we
restrict ourselves to  cylindrical symmetry.

The
density in each flow ``$s$'' is given by the inverse  of the Jacobian of the
transformation so that
\begin{equation}
\rho(\vq_{s})=1/J(\vq_{s}) \, .
\label{e:aat} \end{equation}
The total density  within the caustic is then   given by the summation  over
each flow of each of their densities,
\begin{equation}
\rho(\vx)=\sum_{{\rm flow}\ s}\ \rho(\vq_s).
\label{e:aau} \end{equation}

\subsection{The velocity field, and the generated vorticity} \label{s:vel}

\begin{figure}
\centering \includegraphics[width=5.in]{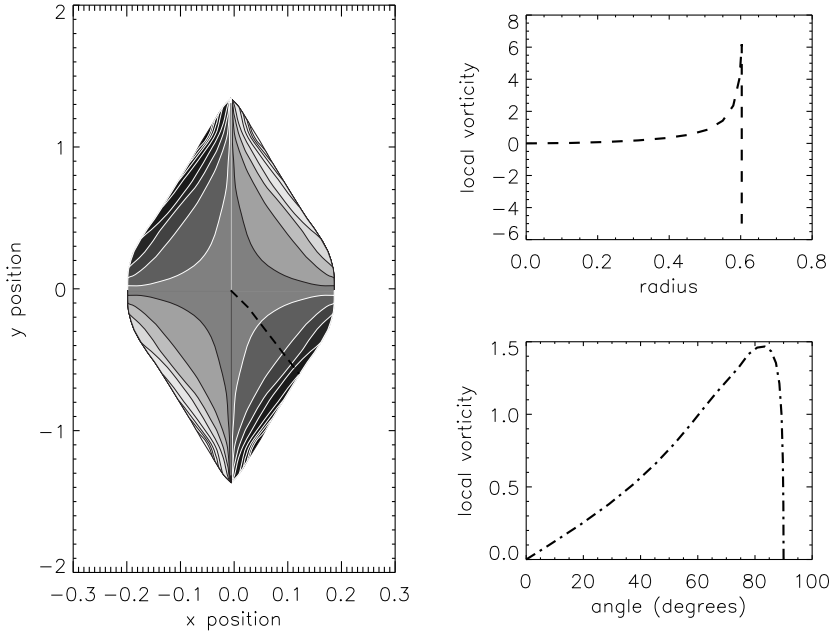}
\caption{The map of the vorticity in a typical 2D caustic ($n=-1$).
Left panel:
the local vorticity  is antisymmetric with respect to the centre of
the caustic. It points along the Z-axis, and is
positive in the second and fourth quadrant, and  negative in the first
and third. 
Right panels: behavior
of the local vorticity along two different lines (thick dot-dashed
line on the left panel). The top panel shows that the vorticity
is singular near the edge of the caustic. It behaves as described by 
Eq. (\ref{e:omegaedge})              
and there is a non zero lineic vorticity located on the edges (represented
here by a vertical line) due to the discontinuity of the local velocity field.
The bottom panel shows that the local vorticity goes continuously to zero
towards the axes.
}
\label{f:fig-7}
\end{figure}

The velocity in each flow is given by
\begin{equation}
\vu(\vq)=\dot{D}(t)/D(t_0)\,\Psi(\vq).
\label{e:aau2} \end{equation}
For a given Robertson Walker cosmology, $\dot{D}(t)$ obeys
\begin{equation}
\dot{D}(t)=f(\Omega)\,H_0\,D(t)\approx\Omega^{0.6}\,H_0\,D(t)  \, .
\label{e:hubble} \end{equation}
where $H_0$ is  the Hubble constant at the  present time and  $f(\Omega)$ is
the  logarithmic derivative  of  the   growing  factor with respect   to the
expansion     factor.  \Eq{hubble} is  the    only place  where the $\Omega$
dependence (and  $\Lambda$  dependence  though    it is  negligible)    will
come into play.  

In   general the velocity  field, $\vu(\vx)$,   is defined as the
density averaged velocities of each flow.  Thus we have
\begin{equation}
\vu(\vx)={\sum_{{\rm flow}\ s}\rho(\vq_s)\,\vu(\vq_s)\over
\sum_{{\rm flow}\ s}\rho(\vq_s)} \, ,
\label{e:aaw} \end{equation}
where the summation is carried on all the flows that have entered the 
neighborhood of $\vx$.
The vorticity is then given by the anti-symmetric derivatives of the 
total velocity with respect to $\vx$:
\begin{eqnarray}
\omega_k(\vx)&=&\sum_{i,j}\,
\epsilon^{k,j,i}\,{\partial\vu_i(\vx)\over\partial\vx_j} 
=\sum_{i,j}\,\epsilon^{k,j,i}\,\left(
\left[\sum_{{\rm flow}\ s}{\partial\rho(\vq_s)\over\partial\vq_{sl}}
(D^{-1})_{j,l}\,\vu_i(\vq_s)\right]\right. 
\left[\sum_{{\rm flow}\ s}\rho(\vq_s)\right]- 
\left[\sum_{{\rm flow}\ s}\rho(\vq_s)\vu_i(\vq_s)\right] \times \nonumber\\&&
\left.\left[\sum_{{\rm flow}\ s}{\partial\rho(\vq_s)\over\partial\vq_{sl}}
(D^{-1})_{j,l}\right]\right)/
\left[\sum_{{\rm flow}\ s}\rho(\vq_s)\vu_i(\vq_s)\right]^2 \, , 
\label{e:aax} \end{eqnarray}
where  $D_{i,j}$ is the matrix of  the transformation between the Lagrangian
space and the Eulerian space,
\begin{equation}
D_{i,j}={\partial\vx_i\over\partial\vq_j}\, ,
\label{e:aay} \end{equation}
and   $\epsilon^{k,j,i}$  the totally antisymmetric   tensor.  The numerical
expression of  the    local   vorticity  follows    from   the roots      of
\Eq{displacement} and the potentials \Eqs{2Du},\Ep{3Du}.

\subsubsection{The local vorticity} \label{s:local}

As illustrated in \Fig{fig-7} (the 2D case) and \Fip{fig-7bis} (the 3D
case), the vorticity is null outside  the caustic. First note that the
vorticity  sign changes from   one quadrant  to another,  so  that the
global vorticity is zero (as it should be), and  note that within each
quadrant the vorticity is rather smooth.  Note also that the vorticity
is mainly  located near   the  edges of  the   caustic.  In fact   the
vorticity at the edge is unbounded  and the behaviour of the vorticity
close to  the edges is easily estimated.   Calling $\vq_0$ and $\vx_0$
the position of a  point on the  edge  in respectively the  Lagrangian
space and the Eulerian  space, we can expand  $\vx$ and $\vq$ close to
$\vx_0$  and  $\vq_0$.   Since the  linear  term  in the  expansion is
singular in $\vq=\vq_0$ (by  definition of the  caustic), there is one
direction,  orthogonal  to the edge    and typeset with  the subscript
$_{\ort}$, for which
\begin{equation}
(\vx-\vx_0)_{\ort}\approx-\eta\,(\vq_i-\vq_0)_{\ort}^2 \, , 
\label{e:vortedge} \end{equation}
where $\eta$   is given by the  second  order expansion of  the displacement
field along this direction.  The minus sign accounts here  for the fact that
$\vx_{0\ort}$  has  been assumed  to  be  larger   than $\vx_{\ort}$.   This
equation is valid for two different flows (say 1 and 2) corresponding to the
two roots of $\vq_i$ in \Eq{vortedge}.  The Jacobian for the first two flows
is then
\begin{equation}
J(\vx)\approx -2\eta\,(\vq_i-\vq_0)_{\ort}\approx
2\,\sqrt{\eta\,(\vx_0-\vx)_{\ort}}.
\label{e:aba} \end{equation}
Note that on the edge  of the caustic, $J(\vx)\vert\partial\,J(\vx)/\partial
\vx\vert$ has a  finite value, $\eta$.  There  is also a  third flow  in the
vicinity of $\vx_0$  which is regular; let us  call $\vq_3$ the Lagrangian
position of $\vx_0$ in this flow.  The velocity is then given by
\begin{eqnarray}
\vu(\vx)\approx\left((\vx_0-\vx)_{\ort}^{-1/2}/\sqrt{\eta}\,
\vu(\vq_0)+\rho(\vq_3)\,\vu(\vq_3)\right)/ 
\left((\vx_0-\vx)_{\ort}^{-1/2}/\sqrt{\eta}+\rho(\vq_3)\right).
\label{e:abb} \end{eqnarray}
As a result we have
\begin{equation}
\vu(\vx)\approx\vu(\vq_0)+\rho(\vq_3)\,\sqrt{\eta}\,(\vx_0-\vx)_{\ort}^{1/2}\,
(\vu(\vq_3)-\vu(\vq_0)),
\label{e:edgev} 
\end{equation}
when $\vx$ is within the caustic and
\begin{equation}
\vu(\vx)=\vu(\vq_3),
\end{equation}
when $\vx$ has crossed the caustic boundary. The local velocity
is thus discontinuous at the caustic boundary and 
 the induced vorticity is consequently singular at  $\vx_0$ with
\begin{equation}
\omega(\vx)\approx
-\rho(\vq_3)\,\sqrt{\eta}\,(\vx_0-\vx)_{\ort}^{-1/2}\,
(\vu(\vq_3)-\vu(\vq_0))_{\para}/2.
\label{e:omegaedge} 
\end{equation}
The direction $\para$ is a direction parallel to the caustic.  There is only
one such direction in 2D, two in 3D.
There is however not only a surface (or volume) contribution within
the caustic. Because of the discontinuity of  the velocity field at
the  edges of the   caustic, a  vorticity field on the boundary
of the caustic is created (see Fig. \ref{f:fig-7} for the 2D case),
whose linear or  surface density  for respectively the  2D and  3D cases are
given by
\begin{equation}
\omega_{\rm lin.,\ surf}=(\vu(\vq_3)-\vu(\vq_0))_{\para}.
\label{e:abd} 
\end{equation}
It turns out that the two contributions tend to cancel  each other.  Indeed,
as we have noticed  previously, the velocity increases close  to the edge of
the caustic, and  then has a discontinuity  at the edge.  This creates a sharp
peak in the vicinity of  the edge of the vorticity.  The vorticity, which is
obtained by differentiation of  the local  velocity  is then expected  to be
opposite on  both   side of this   peak.   Realistically,  the small   scale
perturbations are going to  wash  out these  features,   and to  smooth  the
velocity peaks.  As a result the quantities  describing the behaviour of the
vorticity near the  edge  of the caustic are   not robust and should  not be
taken  at  face value.    On the other   hand,   we expect   the
integrated  vorticity  to be a  more   robust quantity,   since  it is  roughly
independent of  small scale fluctuations.

\subsubsection{The integrated vorticity} \label{s:intvort}
In two dimensions,  the integrated vorticity  in each quadrant can be easily
obtained numerically by simple  one dimensional integrals which, from Stoke's
theorem, can be expressed as
\begin{eqnarray}
\omega_{\rm quad.}=
\int_{\rm quadran}\,\d^2 
\vx\,\omega(\vx)=\int_{\rm edges}\vu\cdot\vdl, \label{e:quad}
\end{eqnarray}
where $\vdl$ describes the  edge of the  quadrant.  One should bear  in mind
that, in  \Eq{quad} the velocities  on the edge  of the caustic are taken as
the velocities of the  third flow, $\vu(\vq_3)$, so  that the singular part of
the    vorticity  is taken into   account.    

In three dimensions and for (almost) spherically symmetric caustics
the local vorticity is independent of the azimuthal angle, $\theta$.
It is then convenient to calculate the integrated vorticity
per azimuthal angle in each quadrant,
\begin{eqnarray}
\omega_{\rm quad.} \, \d \theta =
\left(
\int_{\rm quadran}\,\d z \, r \,  \d r  \,\omega(\vx)
\right) \, \d \theta =\left(
\int_{\rm edges}r \, \vu\cdot\vdl +\int_{\rm quadran}\,\d^2 \vx\, u_z 
\right) \, \d \theta , \label{e:quad3D}
\end{eqnarray}
where $r$ is the  distance of the running  point  to the symmetry axis,  and
$u_z$ is the  velocity component along this axis.   Compared to the 2D  case
there is  a further difficulty due to  the surface integral of one component
of the velocity.  Note nonetheless that this contribution is not singular at
the  edge of the  caustic  as shown by \Eq{edgev},   and can thus be  safely
computed numerically.  We   found  that  this second  integral    contributes
typically to about $15\%$ of the total for the relevant caustics.

\begin{figure}
\centering \includegraphics[width=5.in]{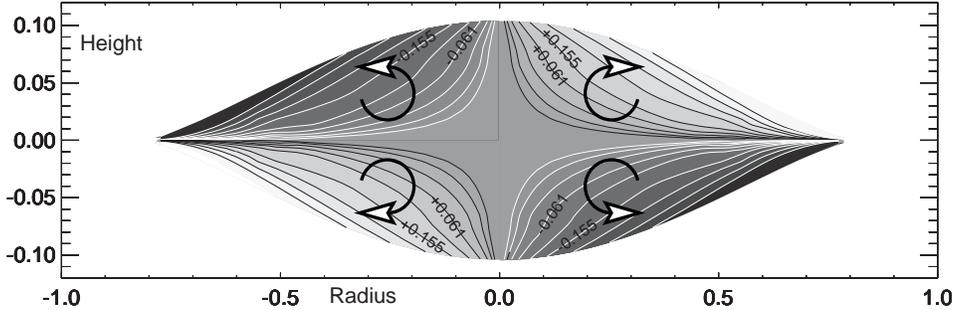}
\caption{Section of the vorticity  field for the caustic of \Fig{fig-6}.
The local vorticity  is antisymmetric with respect to the centre of
the caustic.
In this $X-Z$ section, it points along the Y-axis, and is
negative in the second and fourth quadrant, positive in the first
and third.}
\label{f:fig-7bis}
\end{figure}

\subsubsection{Scaling laws}
\begin{figure}
\begin{minipage}[b]{0.45 \linewidth}
\centering \includegraphics[width=3.in]{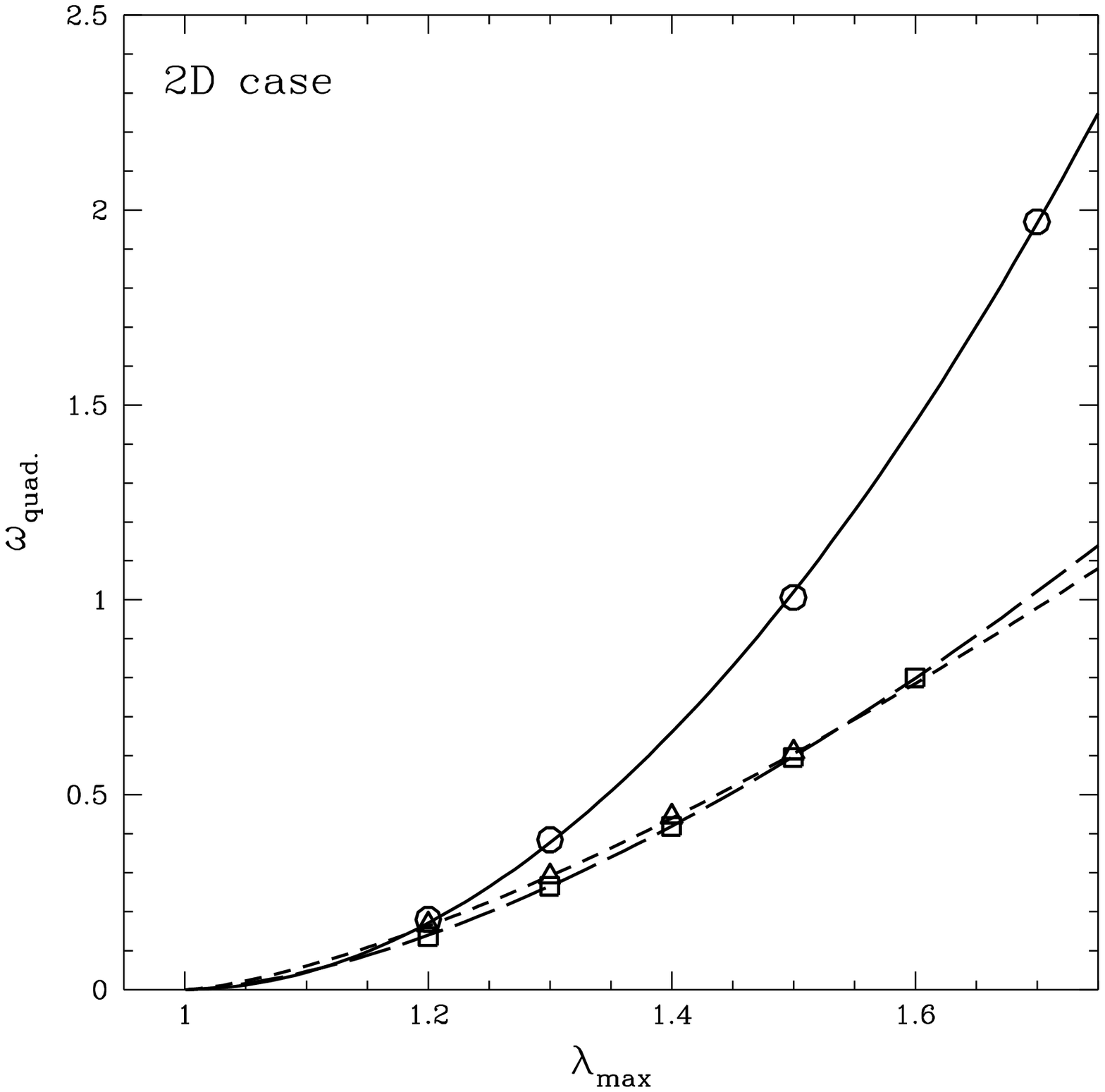}
\caption{$\omega_{\rm quad.}$ for 2D caustics as a function of $\lm$ and its
corresponding  fit for a  $n=-1.5$ (circles,  solid line), $n=-1$  (squares,
long   dash  line), and  $n=-0.5$   (triangles,   short  dashed  line) power
spectrum.}
\label{f:fig-001}
\end{minipage}
\begin{minipage}[b]{0.1 \linewidth}
\end{minipage}
\begin{minipage}[b]{0.45 \linewidth}
\centering \includegraphics[width=3.in]{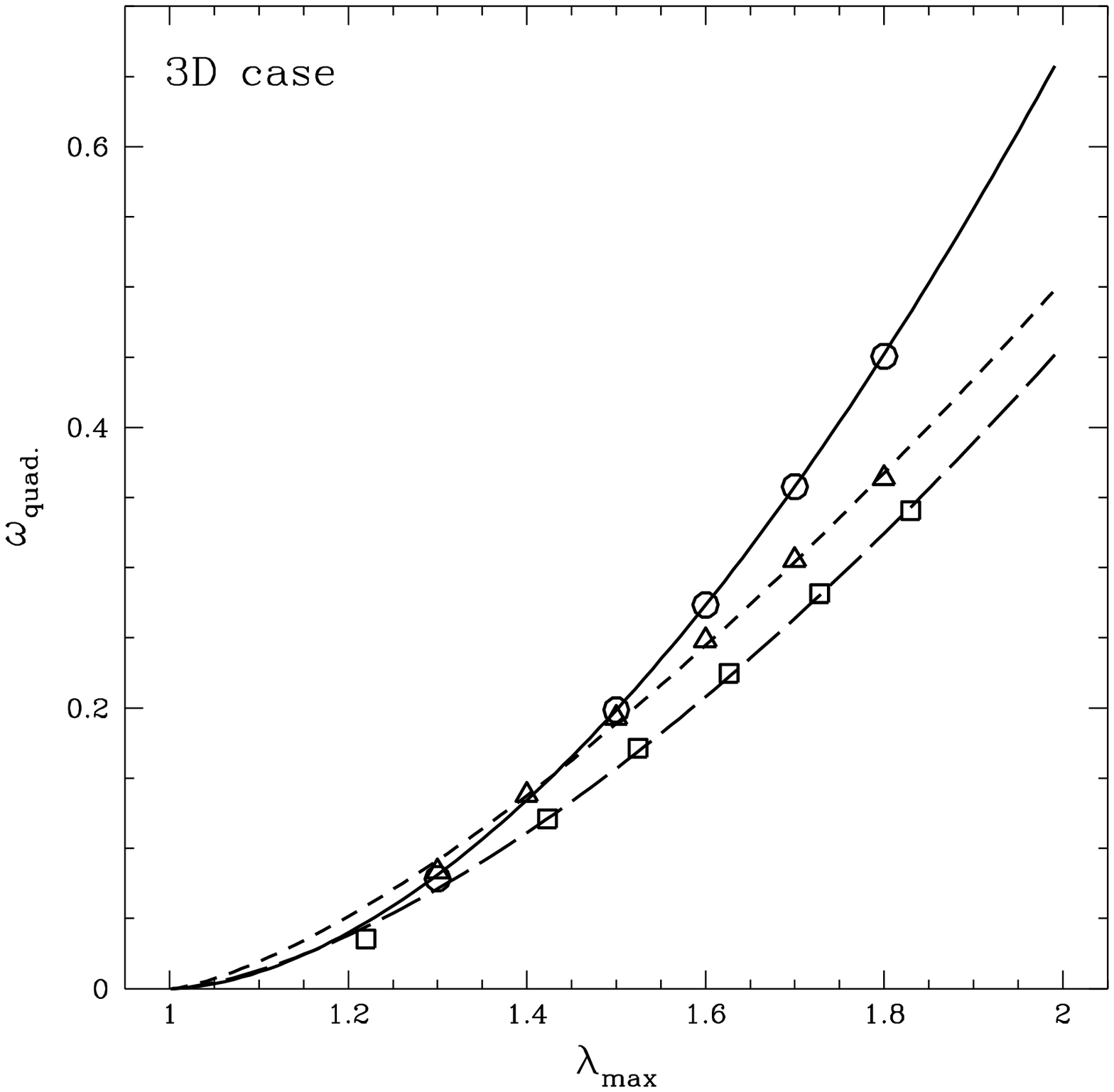}
\caption{$\omega_{\rm quad.}$ for 3D caustics as a function of $\lm$ and its
corresponding fit  for  a $n=-2$  (circles, solid line),  $n=-1.5$ (squares,
long dash line), and $n=-1$ (triangles, short dashed line) power spectrum.}
\label{f:fig-002}
\end{minipage}
\end{figure}

We  now bring forward  fits to   describe the dependence  of the  integrated
vorticity with  the spectral index $n$ and  $\lambda_{\rm max.}$  which will
allow  us to characterize the most  significant  caustics that contribute to
the large--scales vorticity.   We   make explicit the  dependence  of  those
quantities with  respect   to the size   of the  perturbation  $R_L$ and the
cosmological    parameter $\Omega$.   Expressed  in  units  of the expansion
factor, the displacement, in the Zel'dovich approximation, is independent of
$\Omega$.  Therefore $a$ and $b$ are independent of $\Omega$, and are simply
proportional to $R_L$.  The total vorticity in each quadrant is on the other
hand  proportional to $H_0$  and $f(\Omega)$ (defined in \Eq{hubble}), given
that it is  proportional to the local  velocity, and is clearly proportional
to the volume of the perturbation.  We thus have the following scalings,
\begin{equation}
d(R_L)  =  R_L\ d_0\ (\lm-1)^{\alpha_d}, \ \
e(R_L)  =  R_L\ e_0\ (\lm-1)^{\alpha_e},\ \
\omega_{\rm quad} (R_L,\Omega) 
 =  f(\Omega)\ R_L^D\ \omega_0\ (\lm-1)^{\alpha}\,H_0, \label{e:scaling}
\end{equation}
where  the parameters  $\alpha$, $\alpha_{d}$,  $\alpha_{e}$, $ \omega_{0}$,
$d_{0}$ and $ e_{0}$ are given in \Tab{tab1} and \Tap{tab2} for respectively
the  2D and the 3D geometry.   The accuracy of  these fits is illustrated on
\Figs{fig-001}--\Fip{fig-002}.    These functions  yield   estimates of  the
geometry and vorticity generated by these  large-scale caustics.  From these
tables one can see that the average vorticity (in units of $H_0$) is roughly
one within the caustic.   The amount of vorticity  which is generated in the
caustics  is thus found  to be somewhat limited. It   is also interesting to
note that $\omega_{\rm  quad.}$    presents  no singular  behaviour     when
the caustic appears at $\lm\approx 1$ (i.e. $\alpha >1$).

\begin{table}
\caption{Fitting parameters in \Eq{scaling} for the 2D caustics. The quality
of those fits for $\omega_0$ and $\alpha$ are illustrated in \Fig{fig-001}.}
\label{t:tab1}
\begin{tabular}{@{}lcccccc}
$n$&$\omega_0$&$\alpha$&$d_0$&$\alpha_d$&$e_0$&$\alpha_e$\\
-1.5& 3.94 & 1.95 & 0.8 & 1.3 & 2.7 & 0.6 \\
-1& 1.80 & 1.59 & 0.67 & 1.3 & 1.6 & 0.45 \\
-0.5& 1.63 & 1.43 & 0.75 & 1.3 & 1.3 & 0.32
\end{tabular}
\end{table}
\begin{table}
\caption{Fitting parameters in \Eq{scaling} for the 3D caustics. The quality
of those fits are illustrated in \Fig{fig-002}.}
\label{t:tab2}
\begin{tabular}{@{}lcccccc}
$n$ &$\omega_0$&$\alpha$&$d_0$&$\alpha_d$&$e_0$&$\alpha_e$\\
-2  & 0.67 & 1.76 & 0.57 & 1.31 & 1.61 &  0.49 \\
-1.5& 0.46 & 1.55 & 0.52 & 1.30 & 1.25 &  0.37 \\
-1  & 0.49 & 1.37 & 0.53 & 1.30 & 1.13 &  0.30 
\end{tabular}
\end{table}

\section{The vorticity distribution at large scales} \label{s:vortD}

As argued previously, the calculation of the global shape of the  vorticity 
distribution is beyond the scope of this paper.  
Indeed the  low $\omega$ behaviour of the vorticity
distribution is  dominated  by the small  caustics  that are  not rare,  and
therefore not well   described by the  dynamical  evolution of an   isolated
object.  The aim of this section is to estimate {\sl  the shape and position
of the cutoff in the probability distribution function of the local smoothed
vorticity}.     We    will  therefore  estimate    $P_{R_s}(>\omega_s)$, the
probability that in  a circular or  spherical cell of  radius $R_s$ the mean
vorticity exceeds  $\omega_s$.  
This estimation  requires
\begin{itemize}
\item[(i)] identifying the caustics that contribute mostly for each case;
\item[(ii)] estimating the contribution of each of those caustics.
\end{itemize}
In each case  various approximations are  used. In the  main text  we simply
spell the major highlights of the derivation.  A  more detailed and explicit
calculation of the vorticity distribution is presented in Appendix~C.
\subsection{Identification of the caustics}

We  assume   in what  follows  that  $\omega_s$   is large  enough   for the
contribution  to  $P_{R_s}(>\omega_s)$ to  be dominated   by large and  rare
caustics.  This  assumption is the corner stone  of the calculation:  only a
small  fraction   of the caustics   with   specific characteristics  at some
critical time will contribute.

The identification of the caustics contributing most  results of a trade off
between  the   amount of vorticity   a  given caustic  can  generate and its
relative rarity:  the higher $\lm$,  the greater  the internal vorticity is,
according to \Eq{scaling} and given that $\alpha$ is positive, but the rarer
those caustics are  (Equations~\Ep{p2Dlm} and  \Ep{p3Dlm}).  Obviously $\lm$
should be larger than unity for any  vorticity at all to  be generated.  The
calculation is slightly  complicated by the fact  that the Eulerian  size of
the  caustics also depends  of the value of  $\lm$.  Let us assume here that
the  Eulerian  size  of  the  caustics   is substantially smaller  than  the
smoothing length, so that the entire  integrated vorticity in a quadrant can
contribute  (in Appendix~\ref{s:vcaustic},  this assumption  is shown to  be
self-consistent).  This implies a scaling relation between the smoothing cell,
$\omega_s$ and $\lm$,
\begin{equation}
\omega_s\ R_s^D\propto R_L^D\ (\lm-1)^{\alpha}.\label{e:RLRs}
\end{equation}
For a given  smoothing length and  a given $\omega_s$,  \Eq{RLRs} yields  a relation
between the value of $\lm$ and the size  of the caustic.  The caustics which
contribute  most to the vorticity are  then obtained by minimizing the ratio
$\lm^2/\sigma^2(R_L)$    which appears in    the  exponential cutoff of  the
distribution  function  of   $\lm$  (Equations~\Ep{p2Dlm}  and  \Ep{p3Dlm}).
Given that $\sigma^2(R_L)$  behaves like $R_L^{-(n+D)}$ this minimization
yields for the extremum value of $\lm$,
\begin{equation}
\lm^{(0)}={2D\over 2D-\alpha(n+D)}.\label{e:lm0mt}
\end{equation}
Note that for  the values of $\alpha$  we have found,  $\lm^{(0)}$ is always
finite  and positive.   This means that   the  filtered vorticity is  indeed
expected to be dominated  by caustics which have evolved  for a finite time.
This provides an  a posteriori justification  of the  assumptions leading to
this calculation.

The   value  of  $\lm$  found  in  \Eq{lm0mt} is    a robust result  of  our
calculations,  although  it  cannot be  excluded  that this  value could  be
affected  by the  failure of  the Zel'dovich  approximation after the  first
shell crossing.

\subsection{Estimation of the caustic contribution to the vorticity PDF}

In    order to  estimate   the     contribution    of those caustics      to
$P_{R_s}(>\omega_s)$ two other fundamental quantities have to be estimated:
\begin{itemize}
\item[(i)] the number density of caustics;
\item[(ii)] the volume for which each of them contributes to $P_{R_s}(>\omega_s)$.
\end{itemize}
These quantities have been estimated for the specific 
caustics we have previously identified in Sect. 4~1. 

\subsubsection{The number density of caustics}
Estimating the  number density of  caustics  is, in general,  a  complicated
problem.  In the case of Gaussian fields the corresponding investigation was
carried by Bardeen et  al.  (1986) for 3D  fields, and by Bond \& Efstathiou
(1987) for 2D fields.  The  number of caustics is  simply determined by  the
number of points  at which   the first derivatives    of the local   density
vanishes.  This defines accordingly the extrema  of the local density field.
The further requirements we have here on the second order derivatives of the
potential ensures that such  points are in  fact maxima of density field. We
refer here to Bardeen et  al.  (1986) for more details  on how to carry  the
investigation.   A  critical step   involves  transforming the  $\delta_{\rm
Dirac}$ function in  the value of the  first derivatives into a $\delta_{\rm
Dirac}$ function in  the position,  thus  introducing  the  Jacobian of  the
second order derivatives of the density field. After some algebra we find:
\begin{equation}
n_{R_L}(\{\lambda_i\})\,\d^D\lambda_i=         
p\left(\left\{\lambda_i\over
\sigma(R_L)\right\}\right)\,{\d^D\lambda_i\over  \sigma^D(R_L)}
{\vert {\rm Jac}_2(\{\lambda_i\})\vert\over (2\pi\sigma_1^2)^{D/2}} \, ,
\label{e:nRl}
\end{equation}
where the probability $p$ is given either by \Eq{p2D} or \Ep{p3D}
in respectively 2D and 3D, ${\rm Jac}_2(\{\lambda_i\})$ is the Jacobian
of the second order derivatives of the density field for 
given eigenvalues of the deformation matrix and $\sigma_1$
is the variance of the derivatives of the local density
field,
\begin{equation}
\sigma_1^2(R_L)=\int\d^D k\ P(k)\ {\vk^2\over 2}\ W_D^2(R_L).
\label{sig1}
\end{equation}
For a given geometry ({\it i.e.}  given values of $a$ and $b$)
${\rm Jac}_2$ is proportional to $\lm^3$, and it scales as $R_L^{-2\,D}$
due to the derivatives involved in the expression of the matrix
elements. 
It is therefore appropriate to re-express  \Eq{nRl} as
\begin{equation}
n_{R_L}(\{\lambda_i\})\,\d^D\lambda_i=         
p\left(\left\{\lambda_i\over \sigma(R_L)\right\}\right)\,
{\d^D\lambda_i\over  \sigma^D(R_L)}
{n_0(\{\lambda_i\}) \over R_L^D}\,
\left({\lm\over\sigma(R_L)}\right)^D
\quad{\rm where} \quad 
n_0(\{\lambda_i\})=
{\vert {\rm Jac}_2(\{\lambda_i\})\vert \over\lm^D\ (2\pi)^{D/2}}
\ \left[{\sigma\over\sigma_1}\right]^D \, R_L^D.
\label{e:nRl2}
\end{equation}
Note that  $n_0$,  thanks  to the  prefactor  $R_L^D$,  is  a  dimensionless
quantity in  \Eq{nRl2}.  A  further simplification  is provided  by the fact
that  for  large  enough   values  of   $\lm$, the    distribution  function
$p(\{\lambda_i\})$,  at fixed $\lm$, allows  only a  small range of possible
values for the smaller eigenvalues.  We therefore neglect the variations  of
${\rm Jac}_2(\{\lambda_i\})$ with respect  to those variables: it  is viewed
here  as a function of  $\lm$  only and calculated  for  fixed values of the
a-symmetry parameters $a$ and $b$.
The  ratio  $\sigma/\sigma_1$  depends only on  the
value of the power law index. Recall however (see Bardeen et al.  1986) that
this ratio   is not well-defined   for top-hat window  functions  because of
spurious divergences for some values of $n$.  To avoid this problem, we used
the Gaussian window function to compute this ratio.  As  a result, for fixed
values   of $a$ and   $b$, $n_0$ is a dimensionless    quantity that can be
explicitly calculated in a straightforward manner.  Relevant values of $n_0$
are given in tables in the Appendix~\ref{s:vcaustic}.

\subsection{The contributing region}
The region over which  each caustic contributes is  the surface (or volume in
3D) of space in the vicinity of a given caustic where, if one centers a cell
in that location, the total vorticity induced by the caustic within the cell
is above $\omega_s$.

In general the contributing surface or volume can be written,
\begin{equation}
V_{\rm caus.}(R_L,R_s,\{\lambda_i\},\omega_s)=
\int \Theta\left[
\omega_{\bf c }\left({\bf c },R_L,R_s,\{\lambda_i\} \right)- 
\omega_s \right] 
\, \d^D c \, , \label{e:defVc}
\end{equation}
where $\Theta$ is  the Heaviside  step function,  $\bf c$ stands  for the
vector pointing to the center of the  sampling sphere, while $\omega_{\bf
C  }$ is the  vorticity found in that  sphere intersecting  the caustic with
characteristics  $R_s,\{\lambda_i\}$.   In  its   full generality,   $V_{\rm
caus.}$ is a rather complex function of the  scales $R_L$ and $R_s$, and the
eigenvalues   ${\lambda_i}$    through the shape of    the   caustics and of
$\omega_s$.  Yet, since  the functional form of the  rare event tail  in the
probability distribution function  is basically fixed  by the exponential in
\Eq{p2Dlm}, the  only required ingredient for computing $P_{R_s}(>\omega_s)$
is the scaling behaviour of $V_{\rm caus.}$  at its takeoff -- when reaching
the critical time, $\lm^{(0)}$, at which a given caustic  is large enough to
start  contributing.  The   detailed geometry of  the  caustic and its
vorticity   field  accounts only    for a  correction
in a multiplicative factor. 
Consequently  we make approximations describing  the distribution of the
vorticity on the caustic in order to estimate  the scaling properties of
$V_{\rm caus.}$.

\subsubsection{The 2D contributing surface } \label{s:poles}
\begin{figure}
\begin{minipage}[b]{0.45 \linewidth}
\centering \includegraphics[width=3.in]{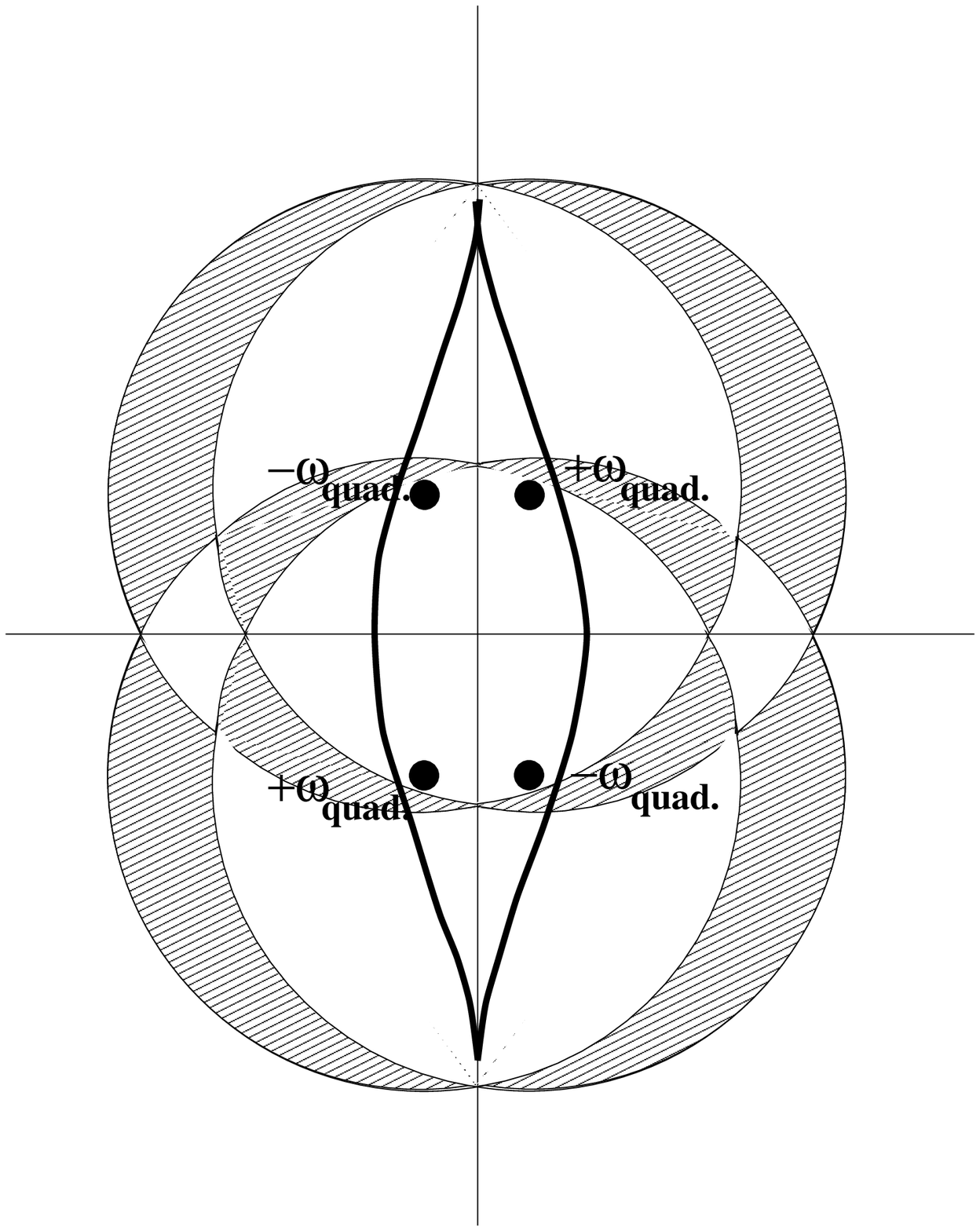}
\caption{Sketch showing the   adopted  simplification for describing a    2D
caustic.  Vorticity is assumed  to be  localized on   the black dots  having
either $+\omega_{\rm quad.}$ or $-\omega_{\rm quad.}$.   The  dashed area
represents $V_{\rm caus.}$ for $\omega_s>\vert\omega_{\rm
quad.}\vert$.
}
\label{f:fig-9}
\end{minipage}
\begin{minipage}[b]{0.1 \linewidth}
\end{minipage}
\begin{minipage}[b]{0.45 \linewidth}
\centering \includegraphics[width=3.in]{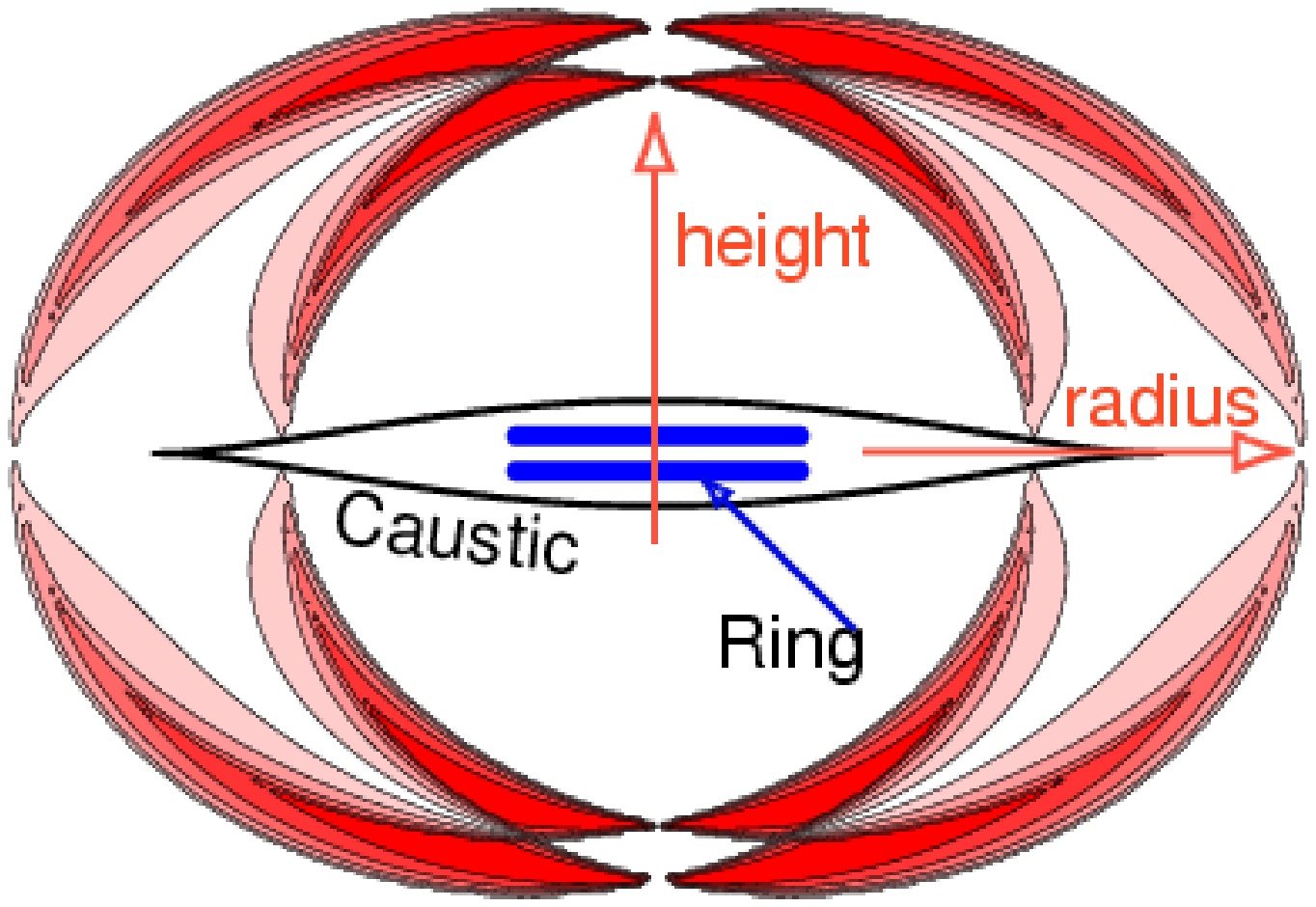}
\caption{Sketch showing the   adopted simplification  for describing a    3D
caustic.  Vorticity  is assumed to be   localized on two rings (that
appear as two horizontal black segments) having
a lineic vorticity of either either  $+3\,\omega_{\rm quad.}/e$ 
or   $-3\,\omega_{\rm quad.}/e$.  The  shaded area
represents $\d V_{\rm caus.}/ \d \omega_s$.
}
\label{f:fig-10}
\end{minipage}
\end{figure}

In two  dimensions we  make the  radical assumption  that  the  vorticity is
entirely concentrated on four discrete points, which  -- consistently with the
hypothesis  of Sect.~\ref{s:intvort}, have  been taken  to bear either the
vorticity $+\omega_{\rm quad.}$ or $-\omega_{\rm quad.}$, depending on which
quadrant is being  considered.  In  practice  the position of  the points is
chosen somewhat arbitrarily at  a third of  the  depth and extension of  the
caustic.  The corresponding  area  $V_{\rm caus.}$ is therefore  identically
null before a  critical time corresponding   to the chosen  $\omega_{s}$ and
$R_s$ and  then takes  a constant value  which can  be deduced geometrically
from   the  area of  the  loci of  the center  of  the   sampling disks.  In
\Fig{fig-9} we show   the shape of  this  location on a  particular example.
Under this assumption, the function $V_{\rm caus.}$  takes the form,
\begin{equation}
V_{\rm caus.}=V_0(R_L/R_s)\ \Theta(\lm-\lm^{(0)})\,R_L\,R_s \, ,
\end{equation}
where $V_0$ can be calculated for the values of interest of $R_L$
and $R_s$.

\subsubsection{The 3D contributing volume } \label{s:rings}
In three   dimensions,  the vorticity   will  be assumed to   be distributed
uniformly along two {\sl rings } which are taken to bear the linear
vorticity   $3\omega_{\rm quad.}/e$   --   with  respectively  prograde  and
retrograde orientation.   In  practice these rings are   also  positioned at a
third of the depth and  extension of the caustic.   The mean vorticity to be
expected in a  sampling sphere of radius $R_s$  is  then given by  algebraic
summation over the segments corresponding to the intersection of that sphere
with the two rings.   Maps of the  sampled vorticity as  a function  of the
centers of the sphere are derived  to compute $V_{\rm caus.}$ which according
to \Eq{defVc} corresponds  to the volume in  space defined by these  centers
yielding a  vorticity larger than  $\omega_s$.  \Fig{fig-10} gives the shape
of this   location for a  given caustic  and sampling radius.   The function
$V_{\rm caus.}$  takes the form,
\begin{equation}
V_{\rm caus.}=V_0(R_L/ R_s) R_L R_s^2 \ (\lm-\lm^{(0)})^{\gamma} \, ,
\label{e:v3dmt}
\end{equation}
where $V_0$ and  $\gamma$ can be calculated  for  the values of interest  of
$R_L$ and $R_s$  at  this critical values (see   Appendix ~\ref{s:asymptote},
where it is in particular demonstrated  that when $R_L  \ll R_s $,
$V_0$ asymptotes to a fixed value and $\gamma =
1/2$).

\subsection{Estimation of $P_{R_s}(>\omega_s)$} \label{s:stat}

The tail of the probability distribution for  the vorticity is now estimated
while integrating over all the  caustics that might contribute, and assuming
that, for  a fixed  caustic, the probability  distribution is  given  by the
number density of caustics  times the volume  associated with each  caustic.
There  is however   a   further difficulty.   The distribution  of  caustics
$n_{R_L}$ is well defined for  a fixed value  of  $R_L$ only, but there  are
actually caustics  of all sizes.  To  circumvent  this difficulty  we simply
choose $R_L$ so that the result we obtain is maximal, i.e.,
\begin{eqnarray}
P_{R_s}(>\omega_s)\simeq 
\max_{R_L}\left[
\int\d^D\lambda_i\ n_{R_L}(\{\lambda_i\})\ 
V_{\rm caus.}(R_L,R_s,\{\lambda_i\},\omega_s)\right]\, .
\label{e:pr1mt}
\end{eqnarray}
Furthermore, it is fair to neglect  the dependence of $n_0({\lambda_i})$ and
$V_{\rm   caus}$ on the    initial   asymmetry because  the  overall  factor
$p({\lambda_i)}$ peaks in a narrow range  of relevant values for the smaller
eigenvalue(s).    It  is then possible   to integrate  over  those variables
introducing  the probability  distribution of $\lm$    in the expression  of
$P_{R_s}(>\omega_s)$,
\begin{eqnarray}
P_{R_s}(>\omega_s)\simeq 
\max_{R_L}\left[
\int{\d\lm}
\ p_{\rm max}\left({\lm}\right)\,
{n_0(\lm) \over R_L^D}\,
\left({\lm\over\sigma(R_L)}\right)^D
V_{\rm caus.}(R_L,R_s,\lm,\omega_s)\right]\, .
\end{eqnarray}

We  show  in  Appendix~\ref{s:vcaustic} that the   maximum of  \Eq{pr1mt} is
indeed given by caustics of size of the order of $R_s$  at most.  A detailed
account of how to perform the sum in \Eq{pr1mt}  is also given there  for the
two geometries.  Repeated use of the rare  event approximation together with
the  geometrical  assumptions  on  the vorticity    distribution sketched  in
Sect.~\ref{s:poles} and Sect.~\ref{s:rings}  yields  eventually an   explicit
expression for the tail of the probability distribution for the vorticity as
a function of $\omega_s$ and $R_s$.

\subsubsection{The two dimensional vorticity distribution}
In two dimensions, the vorticity distribution is shown 
 to obey (\Eq{p2Dfinal})
\begin{equation}
P_{R_s}(>\omega_s)\simeq 
0.56\,n_0\,V_0\,\left({\lm^{(0)}\over \sigma(R_s)}\right)^2\,
f_s^{n+1}\,\omega_s^{(n+1)/2}\,
\exp\left[-{4\over 3}\,\left({\lm^{(0)}\over\sigma(R_s)}\right)^2\,
f_s^{n+2}\,\omega_s^{(n+2)/2}\right].\label{e:p2Dfinalmt}
\end{equation}
In the rare event r\'egime, the quantity that dominates \Eq{p2Dfinalmt} arises
from the exponential cutoff. For $n=-1$ we find for instance that
\begin{equation}
\log\left[P_{R_s}(>\omega_s)\right]\simeq 
3.5\,{\omega_s^{1/2}\over\sigma^2(R_s)}.\label{e:p2Dfinalmtapprox}
\end{equation}
The {\it r.h.s.} of 
\Eq{p2Dfinalmtapprox} is roughly $0.5$ 
   when $\omega_s\approx 10^{-3},\ \sigma(R_s)\approx
0.5$  or $\omega_s\approx    0.1,\  \sigma(R_s)\approx    1.5$,
hence defining   a  threshold corresponding to
 a one sigma damping for $P_{R_s}(>\omega_s)$.
    Equation
\Ep{p2Dfinalmt} is  illustrated on \Fig{fig-f17}.

\subsubsection{The three dimensional vorticity distribution}
\begin{figure}
\begin{minipage}[b]{0.45 \textwidth}
\centering \includegraphics[width=3.5in]{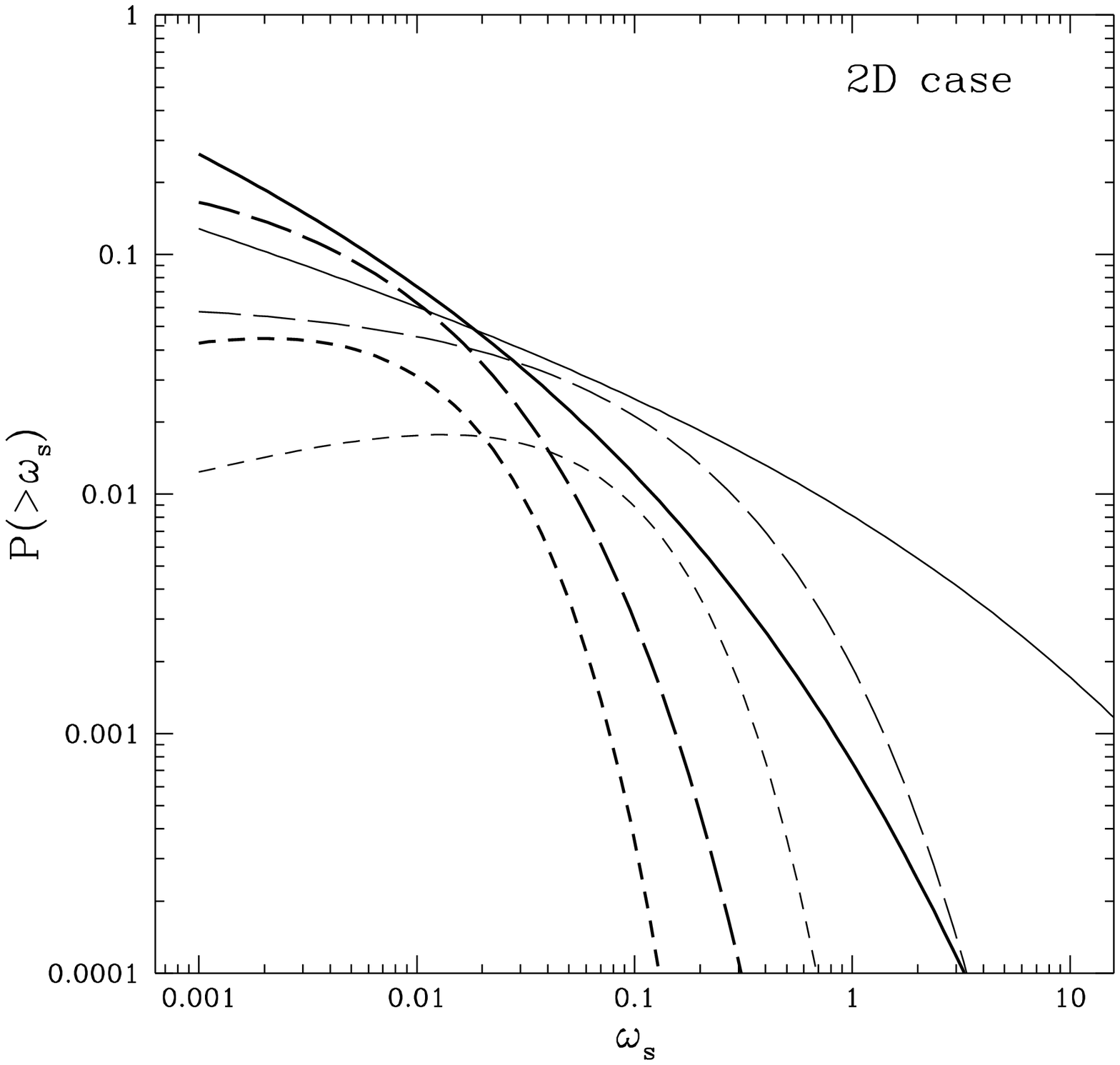}
\caption{ $P_{R_s}(>\omega_s)$ in two dimensions for scales characterized by
a $\sigma(R_s)$ of   $0.5$ (thick lines) and  $1$   (thin lines) and   for a
$n=-1.5$ (solid line),  $n=-1$ (long dash line),  and $n=-0.5$ (short dashed
line) power spectrum.}
\label{f:fig-f17}
\end{minipage}
\begin{minipage}[b]{0.1 \textwidth}
\end{minipage}
\begin{minipage}[b]{0.45 \textwidth}
\centering \includegraphics[width=3.5in]{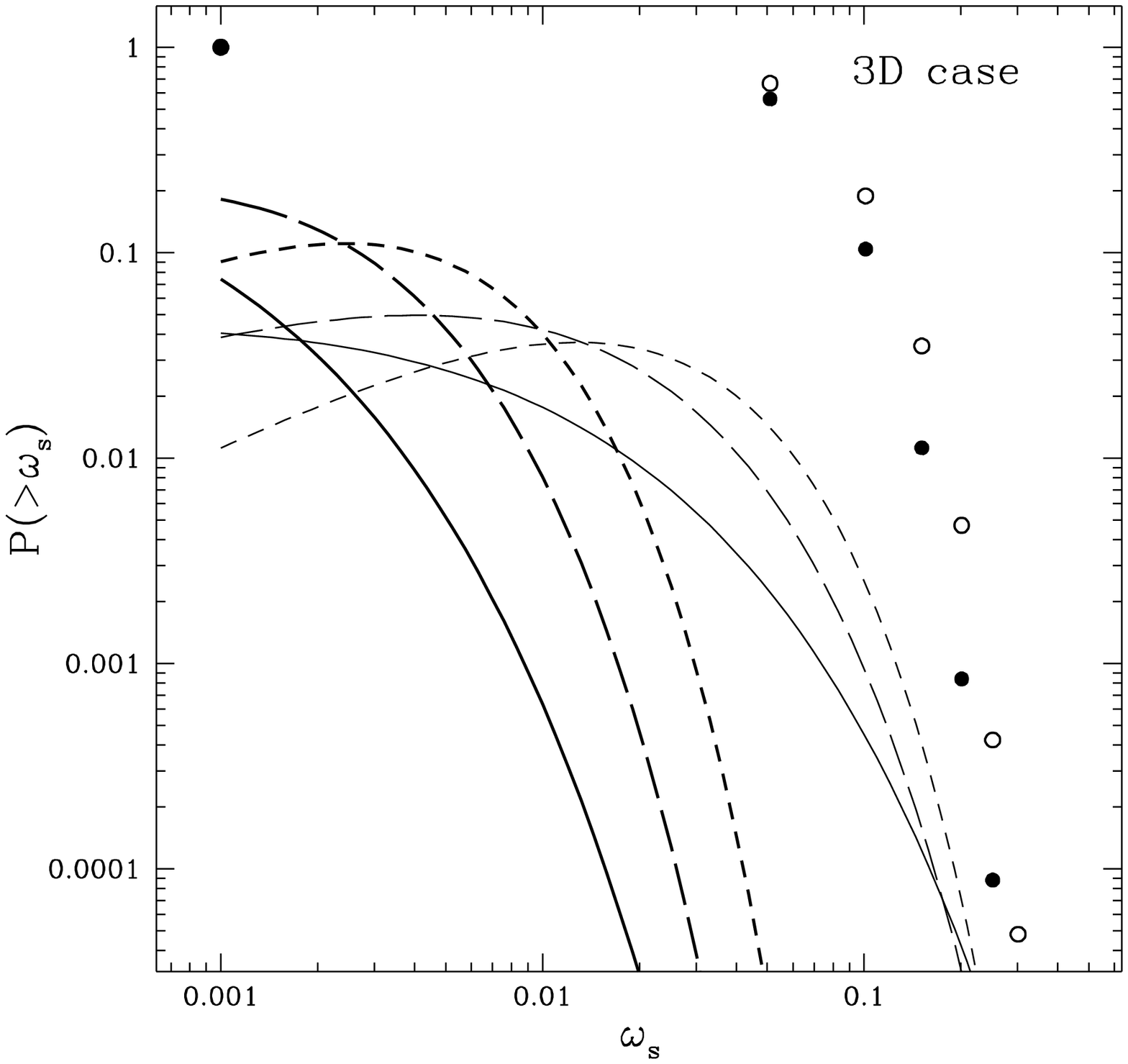}
\caption{ $P_{R_s}(>\omega_s)$ in three  dimensions for scales characterized
by a $\sigma(R_s)$  of $0.5$ (thick lines)  and $1$ (thin  lines)  and for a
$n=-2$ (solid  line), $n=-1.5$  (long dash  line), and $n=-1$  (short dashed
line) power spectrum.  The  filled and open circles correspond  respectively
to the  measured integrated PDF in a  CDM  simulation at $15h^{-1}$Mpc scale
with the  ``Delaunay''  or ``Voronoi''  methods (see  Bernardeau  \& Van  de
Weygaert 1996).}
\label{f:fig-f18}
\end{minipage}
\end{figure}

Similarly,    the     probability    distribution   is     shown  in     the
Appendix~\ref{s:vcaustic} (\Eq{p3Dfinal}) to obey in 3D:
\begin{equation}
P_{R_s}(>\omega_s) = 0.48\,n_0\,V_0\,
\left({\lm^{(0)}\over \sigma(R_s)}\right)^{7/2}\,
f_s^{(13+7n)\over 4}\,\omega_s^{(13 +7 n) \over 12}\,
\exp\left[-{5\over 2}\,\left({\lm^{(0)}\over\sigma(R_s)}\right)^2\,
f_s^{n+3}\,\omega_s^{(n+3)/3}\right], \label{e:p3Dfinalmt}
\end{equation}
 For $n=-1.5$, \Eq{p3Dfinalmt}
gives for $\log\left[P_{R_s}(>\omega_s)\right]$
\begin{equation}
\log\left[P_{R_s}(>\omega_s)\right]\simeq 
20.\,{\omega_s^{1/3}\over\sigma^2(R_s)}.
\end{equation}
yielding  again  at a  one  sigma level  the   range of relevant  values for
$\omega_s$      and        $\sigma(R_s)$:   $\omega_s\approx    5\,10^{-5},\
\sigma(R_s)\approx  0.5$  or $\omega_s\approx 0.1,\ \sigma(R_s)\approx 3.5$.
In both cases the caustics  start to generate  significant vorticity only at
rather small scales.   Equation   \Ep{p3Dfinalmt}  is also  illustrated   on
\Fig{fig-f18}.
From this figure it is clear that the amount of vorticity that we derived is
below what  has  been measured  in   $N$-body simulations (open  and  filled
circles).  Numerical measurements of this  quantity are  sparse, so we
compared our  estimations  to measurements carried out by  Bernardeau  \& Van de
Weygaert (1996) in an adaptive P$^3$M simulation with CDM initial conditions
(see  Couchman 1991  for a description  of these  simulations).  The typical
amount of vorticity at the $10$ to $15$  $h^{-1}$Mpc scale for which the rms
of the density is below $0.5$ was  found to  be   about $0.2$ (in units of
$H_0$). This is  well  above the values   we have estimated in  this  paper.
Though  it is quite possible  that these numerical measurements are spoiled
by  noise, we do  not expect that it could  account for  all the discrepancy
between the  measured and  the  predicted vorticities  (as suggested by  the
relative the scatter between the two methods  suggested in Bernardeau \& Van
de Weygaert, 1996).

There are various possible explanations for such discrepancies.  It could of
course arise from the fact that the vorticity at large-scales does not spring
from the rare  and large caustics  but from small scale multi-steaming events
that cascade  towards the larger scales.   Such a scenario cannot be excluded
but  is  difficult to investigate  by  means of analytic calculations.
It is  also   possible  that  the $N$-body  simulations do  not address
properly the physics of the large scales  multi-streaming.  In particular the
two-body  interactions should in principle  be  negligible, a property which
seems  to be   hardly   satisfied in current  $N$-body   simulations.   This
shortcoming has  been raised by Suisalu  \& Saar  (1995), Steinmetz \& White
(1997)  and more specifically by Splinter  et  al.  (1998), where they examine
the outcome  of the planar  singularity in phase  space.  They have found in
particular that in classical algorithms  the particle's velocity  dispersions
are incorrectly large in all directions.  These could turn out to be a major
unphysical source of vorticity (since the Lagrangian  time derivative of the
vorticity    scales like  the curl   of   the divergence   of  the  velocity
anisotropies).  Specific numerical experiments, that follow for instance the
initial density  profiles given in this paper,  should be carried to address
this problem more carefully.

\section{Discussion and Conclusions} \label{s:concl}

We have estimated,  within  the framework  of  the gravitational instability
scenario, the amount of vorticity generated after  the first shell crossings
in  large-scales    caustics.  The calculations  relied   on  the Zel'dovich
approximation  which yields estimates of  the characteristics of the largest
caustics and allows  explicit calculation of their  vorticity content.  This
analysis  corresponds   to one of  the   first  attempts to  investigate the
properties     of   cosmological  density      perturbations  beyond   first
shell-crossing.  The  previous investigations  (Fillmore \& Goldreich  1984,
Bertschinger 1985) were carried out for  spherically symmetric systems only,
and obviously do not address the  physics of vorticity generation.  The only
other  means of   investigation for this   r\'egime  is  numerical $N$-body
simulations.

We  found  that large  scales caustics   can provide  only  an extremely low
contribution  to the vorticity  at scales of   $10$ to $15 h^{-1}{\rm Mpc}$.
This contribution could be significant only at relatively small scales, when
the   variance reaches values  of  a few  units.   This  effect is even more
important   in  three dimensions, the  difference    arising mainly from the
coefficient in the exponential cut-off.  It is therefore unlikely that these
caustics can have  produced a significant   effect on the  velocity at large
scales.  In view of these results, it is amply  justified to assume that the
velocity remains potential down to very small scales,  {\it i.e.}  typically
the cluster  scale   at which   it is   then  more natural    to expect  the
multi-streaming r\'egime (not only three-flow r\'egime) to play an important
role.  

This result provides a complementary view to the picture developed by
Doroshkevich (1970) describing the emergence of galaxy angular
momentum  from  small-scale  torque  interactions    between
protogalaxies   (a prediction   subsequently  checked by   White (1984), and
examined in more detail by  Catelan \& Theuns, (1996 and 1997)). We 
rather explore here 
the large scale coherence of the vorticity field that may emerge
in a hierarchical scenario from scale much larger than the galactic
size. The effects we are exploring here does not 
originate from the two-body interaction of haloes as in the
picture developed by Doroshkevich, but from the possible
existence of large scale coherent vorticity field.
The conclusion of our work is however that the
efficiency with which the large-scale structure caustics
generate vorticity is rather low. Therefore these results
do not really challenge the fact 
that  the  small scales     interactions  should indeed   be  the   dominant
contribution to the actual galactic angular momenta.

As a consequence, we do not expect either a significant  correlation of the
angular momenta at large scale. In particular the vorticity
field generated in caustics does not seem
to be able to induce a significant large scale correlation of the galactic
shapes which would have been desastruous for weak lensing 
measurements\footnote{In these measurements background galaxy shapes are
assumed to be totally uncorrelated in the source plane, the observed
correlation being interpreted as entirely due to gravitational lens effects.}.

Let us reframe this calculation in  the context of perturbation theory which
has triggered some interest  in the last few  years as a tool to investigate
the quasi-linear    growth of  structures.    One  key assumption   in these
techniques is that the velocity field is  assumed to form a single potential
flow.  The detailed description of the properties of the first singularities
is by essence not accessible  to this theory:  such singularities cannot  be
``seen'' through  Taylor expansions of the initial  fields.  In this context
it was unclear whether the back reaction of the small scales multi-streaming
r\'egime on the larger scales (which were thought to be adequately described
by perturbation  theory) could affect  the results  on those  scales.   Such
effects  are partially explored  here where we  find that the  impact of the
first multistreaming  regions is rather low on  larger  scales.  Our results
therefore support  the  idea that the large   scales velocity field  can  be
accurately   described by  potential  flows  and support  our  views  on the
validity domain of perturbation theory calculations.

In the  course  of this derivation  we have  made  various assumptions.   We
followed in essence the approach pioneered by Press  \& Schechter (1974) for
the mass distribution of  virialized objects by  trying  to identify in  the
initial density field the density fluctuations that contribute mostly to the
large-scales    vorticity.   The calculations  have  been  designed  to be as
accurate as possible  in the rare  event limit, an approximation which turned
out to be crucial at various stages of the argument.
\begin{itemize}
\item
The above estimation relies heavily on the assumption that the caustics only
contribute to   large-scale vorticity independently of   each other.  In
other words it is  assumed that the caustics  do not overlap.   Moreover the
dynamical evolution  of one  caustic is taken  to  be well-described by  the
evolution of the caustic having the mean profile.  This can be approximately
true  only in the   rare event limit since otherwise   it is likely that the
substructures and its environment will change the dynamical evolution of the
caustics.   Although it is clear that,  in the r\'egime we investigated, the
caustics  are rare enough  not to overlap, the  effects  of substructure are
more difficult  to investigate.  In  particular we have outlined  some local
features (\ref{s:local}) of the vorticity maps that we think are unlikely to
survive the existence of substructures.
\item
The  typical  caustics are  characterized  in  this  rare  event  limit. For
instance the values of $a$ and  $b$ were found to  be all the more peaked to
given values as the  corresponding events are  rare.  We have then estimated
the vorticity such caustics  generate while assuming that slightly different
geometries are unlikely to produce  very different results.  This assumption
is somewhat  suspicious,  since it might   turn out that  slightly different
geometries  could  produce  more  vorticities,  and thus   change  the exact
position of the cut-off.  We do not expect  however that the conclusions  we
have reached could be changed drastically in this manner.
\item
The  contributions of each  caustics to $P_{R_s}(>\omega_s)$  have also been
calculated in  the rare event   limit. This is  in   practice a very  useful
approximation on large scales since it is then  natural to expect the entire
distribution to be dominated by a unique value of $\lm$.
\item
 We have finally deliberately simplified the spatial distribution of  
the vorticity within the caustics.  Since in the rare event
limit it is natural to  expect that the  Lagrangian scales of the caustics are
much smaller than  the smoothing scale this  detailed arrangement should be
of little relevance.  It certainly should  not affect the scaling laws as only
the value   of  the  overall  factor    $V_0$  will change, 
and this has little bearing on our conclusions.
\end{itemize}

On top of the rare  event limit approximation, we have  also made a dramatic
simplification by using   the Zel'dovich approximation throughout.  This  is
certainly a   secure assumption before  the  first shell-crossing  since the
geometries that we have  investigated were rather sheet-like structures (and
the Zel'dovich   approximation is exact  in 1D  dynamics).   After the first
shell-crossing  however, the back  reaction of the large over-densities that
are created  could possibly affect  the velocity  field. However we  do  not
expect that this effect should be very large so  long as $\lm$ is moderately
small (up to  about 1.5), since before   then the initial  inertial movement
should dominate.  Later on, matter is expected to bounce  back to the center
of the caustics.  Whether the vorticity content is then amplified or reduced
remains an open question.

\vskip 1cm
{\sl CP  wishes to thank J.F.~Sygnet, D.~Pogosyan, S.
Colombi and  J.R.~Bond for useful
conversations.   
Funding from the Swiss  NF  is gratefully acknowledged.  }


\appendix
\section{ Average profile of an a-spherical constrained random field}
\label{s:avg}
\subsection{General Formula}
Let   us evaluate here  the average  profile  of  an a-spherical constrained
random field in both 2 and  3D.  Similar calculations  as those presented in
this Appendix  have been  investigated by Bardeen  et al.  (1986) for the 3D
field and by Bond \& Efstathiou (1987) for 2D fields.  But, here, instead of
the second  order derivative of  the density field,  we consider instead the
deformation  tensor   corresponding   to second order   derivatives   of the
potential.  We also investigate  the global properties that such constraints
induce on the density field.

Consider a  random  density field,  in either 2D  or 3D, having fluctuations
following a  Gaussian  statistics.  It is   then entirely determined,  in  a
statistical sense,  by the shape of its  power spectrum, $P(k)$. Recall that
$P(k)$ is defined from the Fourier transform of the density field,
\begin{equation}
\delta(\vk)=\int\d^3\vx \exp(\i \vk.\vx)\ \delta(\vx) \, ,
\ \ {\rm with}\ \
\mg\delta(\vk)\,\delta(\vk')\md=\delta_{\rm Dirac}(\vk+\vk')\,P(k),
\label{e:abf} \end{equation}
where the brackets $\mg.\md$  stands for the ensemble  average of the random
variables.  Let  us calculate the  {\sl  expectation} value of $\delta(\vk)$
when a local constraint has been set in order to create an {\sl a-spherical}
perturbation.   To set such a  constraint,  we have  chosen to consider  the
deformation  tensor of  the density    field  smoothed  at  a  given   scale
$R_L$. This tensor reads,
\begin{equation}
\phi_{i,j}=\int\d^3\vk\ \delta(\vk)\ W_D(k\,R_L)\ {\vk_i\vk_j\over k^2}.
\label{e:abg} \end{equation}
Note that   the  local smoothed  density   is given  by   the  trace of this
tensor. The  chosen window function $W_D$ in  Fourier space corresponds to a
top-hat filter in real space and it reads,
\be
W_2(k)=2\ {J_1(k) \over k^{1/2}}\quad {\rm in\ }\ \ 2D,\ \ 
W_3(k)=3\sqrt{\pi/2}
\ {J_{3/2}(k) \over k^{3/2}}\ \ {\rm in\ }\ \ 3D,\label{e:abh} 
\end{equation}
where $J_{\nu}$  are the Bessel   functions of   index  $\nu$.  The   matrix
$\phi_{i,j}$ is now set to be equal to a given constraint.  It is obviously
possible  to choose the  axis so that this constraint  is  a diagonal matrix
with   eigenvalues    $(\lambda_i), i=1,D$.  The   elements   of  the matrix
$\phi_{i,j}$ and $\delta(\vk)$ form a {\sl Gaussian} random vector,
\begin{equation}\begin{array}{cl}
V_c=\left(\delta(\vk),\phi_{1,1},
\dots,\phi_{D,D},\phi_{1,2},\dots,\phi_{1,D},\right.
\left.\phi_{2,2},\dots,\phi_{D,D-1}\right),
\end{array}\label{e:abi} \end{equation}
and the desired  expectation value of  $\delta(\vk)$ is  directly related to
the cross-correlation matrix of the components of this vector.
Consider the matrix $C_{a,b}$ with $a=0,  \cdots D (D+1)/2$ and $b=0,
\cdots D(D+1)/2$, so that
\begin{eqnarray}
C_{0,0}&=&\mg\delta(\vk)\,\delta(\vk)\md=P(k) \, , \\
C_{a,0}&=&\mg\delta(\vk)\,\phi_{i,j}\md  
=P(k)\,W_D(k\,R_L)\,{\vk_i\vk_j\over k^2} \,,  \\
C_{a,b}&=&\mg\phi_{i,j}\,\phi_{i',j'}\md 
=\int\d^3\vk\ P(k)\,W_D^2(k\,R_L){\vk_i\vk_j\vk_{i'}\vk_{j'}\over k^4} \, , 
\label{e:abk} \end{eqnarray}
where the indices ${i,j}$  (respectively ${i',j'}$) for the  matrix elements
$\phi_{ij}$ corresponds to  the  $(a+1)^{\rm th}$ (respectively  $(b+1)^{\rm
th}$) component of $V_c$.
For a given spectrum these quantities are easily calculated and are given in
the  following   subsections for  power  law  spectrum  in    resp. 2  and 3
dimensions.  The distribution function of the  components of the vector $V_c$
then reads in terms of \Eq{abk}, 
\begin{equation}
p(V_c)\,\d V_c=
\exp\left[-{1\over 2}
\sum_{a,b}\,\left(C^{-1}\right)_{a,b}\,Vc_a\,Vc_b\right]
\,{\d V_c\over [2\pi \Det(C)]^{1/2+D(D+1)/4}}.
\label{e:abl} \end{equation}
The expectation value of $\delta(\vk)$ is  given by the ratio
\begin{equation}
\delta^{\rm expec.}(\vk)=
{\int\d\delta(\vk)\ \delta(\vk)\ p(Vc)\over
\int\d\delta(\vk)\ p(Vc)} \, ,
\label{e:abm} \end{equation}
A straightforward calculation shows that this quantity is given by
\begin{equation}
\delta^{\rm expec.}(\vk)=
\sum_{i=1}^D-{\left(C^{-1}\right)_{0,i}\over\left(C^{-1}\right)_{0,0}}
\ \lambda_i \, .
\label{e:Cgen} \end{equation}
Note that the further constraint that 
the first derivative of the density field should be zero  (so that 
the point $x_0$ is actually located on a maximum of the density
field) would not change the resulting expression of
$\delta^{\rm expec.}(\vk)$
since the cross correlation of the first order derivatives
with any other involved quantities identically vanish.

\subsection{The 2D profile}
\noindent
In 2 dimensions we have
\begin{equation}
C_{a,b}=
\left(
\begin{array}{cccc}
C_{0,0}&C_{0,1}&C_{0,2}&C_{0,3}\\
C_{0,1}&3\sigma_0^2/8&\sigma_0^2/8&0\\
C_{0,2}&\sigma_0^2/8&3\sigma_0^2/8&0\\
C_{0,2}&0&0&\sigma_0^2/8 
\end{array}
\right) \, ,
\label{e:C2D} \end{equation}
with the variance of the smoothed density field, $\sigma_0$, 
given by
\begin{equation}
\sigma_0^2=\int\d^3\vk\ P(k)\ W^2_D(k\,R_L).
\label{e:abo} \end{equation}
The required elements of the inverse of this matrix are given by
\begin{eqnarray}
\left(C^{-1}\right)_{0,0}&=&
{1\over 64}\sigma_0^6/\Det(C) \, ,\\
\left(C^{-1}\right)_{0,1}&=&
-\left|
\begin{array}{ccc}
C_{0,1}&C_{0,2}&C_{0,3}\\
\sigma_0^2/8&3\sigma_0^2/8&0\\
0&0&\sigma_0^2/8
\end{array}
\right|\ 
{1\over 64\ \Det(C)} 
={\left(C_{0,2}-3\,C_{0,1}\right)\sigma_0^4\over 64\ \Det(C)} \, ,\\
\left(C^{-1}\right)_{0,2}&=&
\left|
\begin{array}{ccc}
C_{0,1}&C_{0,2}&C_{0,3}\\
\sigma_0^2/8&\sigma_0^2/8&0\\
0&0&\sigma_0^2/8
\end{array}
\right|\ 
{1\over 64\ \Det(C)}
={\left(C_{0,1}-3\,C_{0,2}\right)\sigma_0^4\over 64\ \Det(C) } \, .
\label{e:abp} \end{eqnarray}
As a result, \Eq{Cgen} becomes here
\begin{equation}
\delta^{\rm expec.}(\vk)={P(k)\,W_2(k\,R_L)\over\sigma_0^2} 
\left(\lambda_1+\lambda_2+2 \cos(2\theta)[\lambda_1-\lambda_2]\right)
\, ,
\label{e:Cpolar2D} \end{equation}
where the angle $\theta$ were chosen  so that
\begin{eqnarray}
k_1/k=\cos(\theta) \, , \quad 
k_2/k=\sin(\theta).\nonumber
\label{e:abr} \end{eqnarray}
$\theta$ the angle  between  a given vector and  the eigenvector
associated to the first eigenvalue.

\subsection{The 3D profile}
\noindent
In 3 dimensions   the matrix $C$ reads,
\begin{equation}
C=\left(
\begin{array}{ccc}
C_{0,0}&\dots&C_{0,6}\\
\vdots&D&\\
C_{0,6}&&
\end{array}
\right) \, ,  \quad {\rm with } \quad
D={\sigma_0^2\over 15}
\left(
\begin{array}{cccccc}
3&1&1& & & \\
1&3&1& &0& \\
1&1&3& & & \\
 & & &1&0&0\\
 &0& &0&1&0\\
 & & &0&0&1\\
\end{array}
\right).
\label{e:abt} \end{equation}
From this expression of the matrix of the cross correlations
it is quite straightforward to re-express \Eq{Cgen} as
\begin{equation}
\delta^{\rm expec.}(\vk)=
{3 P(k)\,W_3(k\,R_L)\over 2}
\left(\lambda_1[\vk_2^2+\vk_3^2-4\vk_1^2]+\right. 
\left.\lambda_2[\vk_1^2+\vk_3^2-4\vk_2^2]+
\lambda_3[\vk_1^2+\vk_2^2-4\vk_3^2]\right).
\label{e:Cgen3D} \end{equation}
When the coordinates of the wave vector are expressed 
in terms of the angles $\theta_k$ and $\phi_k$, 
defined by
\begin{eqnarray*}
k_1=k\,\sin(\theta_k)\ \cos(\phi_k) \quad
k_2=k\,\sin(\theta_k)\ \sin(\phi_k)\quad {\rm and} \quad
k_3=k\,\cos(\theta_k)\, .
\label{e:abt2} \end{eqnarray*}
\Eq{Cgen3D} becomes 
\begin{eqnarray}
\delta^{\rm expec.}(\vk)=
{3\,P(k)\,W_3(k\,R_L)\over 8\,\sigma_0^2}\,(\lambda_1+\lambda_2+6\lambda_3)
\left(
1+a\,\cos(2 \theta_k)+b\,\cos(2\phi_k)[1+\cos(2 \theta_k)]\right) \, ,
\label{e:Cgen3Dpolar} \end{eqnarray}
where $a$ and $b$ are specific combinations of the eigenvalues,
\begin{equation}
a=5\,
{2\lambda_3-\lambda_1-\lambda_2\over \lambda_1+\lambda_2+6
\lambda_3}\, ,  \quad
{\rm and } \quad
b=5\,{\lambda_1-\lambda_2\over \lambda_1+\lambda_2+6\lambda_3}.
\label{e:abv} \end{equation}

\section{ The DF of the eigenvalues of the local deformation tensor}
\label{s:df}
The derivation of the distribution function of  the eigenvalues of the local
deformation tensor was carried in 3D by Doroshkevich (1970).  We extend here
the calculation   to   the 2D    case   (for which  the    calculations  are
straightforward).     Starting   with    equation    \Ep{C2D}  --     the
cross-correlations between the elements of the  deformation tensors, one can
easily  get  the  expression of   the  joint distribution   function  of  the
deformation tensor elements,
\begin{equation}
p(\phi_{1,1},\phi_{1,2},\phi_{2,2})\ 
\d\phi_{1,1}\ \d\phi_{1,2}\ \d\phi_{2,2}=
{8\over (2 \pi)^{3/2}}\ 
{\d\phi_{1,1}\ \d\phi_{1,2}\ \d\phi_{2,2}\over \sigma_0^3} \
\exp\left[-{1\over2}
\left(3 \phi_{1,1}^2+8\phi_{1,2}^2+3\phi_{2,2}^2-2\phi_{1,2}\phi_{2,2}\right)
\right]
\label{e:abw} \end{equation}
The change of variables,
\begin{equation}
\lambda_+={\phi_{1,1}+\phi_{2,2}\over 2}
+{\sqrt{\Delta}\over2} \, ,\quad
\lambda_-={\phi_{1,1}+\phi_{2,2}\over 2}
-{\sqrt{\Delta}\over2}\, ,
\quad {\rm with} \quad
\Delta
=(\phi_{1,1}-\phi_{2,2})^2+4\phi_{1,2}^2,
\label{e:aby} \end{equation}
allows us to introduce the eigenvalues of the matrix.
The Jacobian $J$ of this transformation is given by
\begin{equation}
\begin{array}{cl}
J^{-1}=
\left\vert
\begin{array}{ccc}
{1\over2}+{\phi_{1,1}-\phi_{2,2}\over2\sqrt{\Delta}}&\sim&
{1\over2}-{\phi_{1,1}-\phi_{2,2}\over2\sqrt{\Delta}}\\
\\
{1\over2}-{\phi_{1,1}-\phi_{2,2}\over2\sqrt{\Delta}}&\sim&
{1\over2}-{\phi_{1,1}+\phi_{2,2}\over2\sqrt{\Delta}}\\
\\
0&1&0
\end{array}
\right\vert 
=\sqrt{1-4\,\phi_{1,2}/\Delta}
\end{array} \, .
\label{e:aby2} \end{equation}
As a result we have 
\begin{equation}
p(\lambda_+,\lambda_-,\phi_{1,2})\ 
\d\lambda_+\ \d\lambda_-\ \d\phi_{1,2}= 
{8\d\lambda_+\ \d\lambda_-\ \d\phi_{1,2}\over (2\pi)^{3/2}\sigma_0^3}
\ {1\over\sqrt{1-4\,\phi_{1,2}/\Delta}}\ 
\exp\left[-{1\over\sigma_0^2}\big({3\over2}J_1^2-4J_2\big)\right]
\nonumber \, ,
\label{e:abz} \end{equation}
with
\begin{eqnarray}
J_1=\lambda_++\lambda_-\ , , \quad {\rm and } \quad
J_2=\lambda_+\ \lambda_-.
\label{e:aca} \end{eqnarray}
The integration over $\phi_{1,2}$ yields 
\begin{equation}\begin{array}{l}
p(\lambda_+,\lambda_-)\ 
\d\lambda_+\ \d\lambda_-= 
\sqrt{2\over\pi}\ {\d\lambda_+\ \d\lambda_-\over\sigma_0^3}
\vert\lambda_+-\lambda_-\vert
\ \exp\left[-{1\over\sigma_0^2}\big({3\over2}J_1^2-4J_2\big)\right].
\end{array}\label{e:acb} \end{equation}
Note that if $\lambda_+$ is a priori assumed to be greater
than $\lambda_-$ the distribution should be multiplied by 2.

\section{Estimation of $P_{R_s}(>\omega_s)$} \label{apc}
\label{s:vcaustic}

In    this   Appendix  we  estimate     the  probability
$P_{R_s}(>\omega_s)$  that a sphere  of radius  $R_s$ contains an integrated
vorticity larger than $\omega_s$.  
In  order to account for  caustics of all sizes  we argued  in the main text
that $P_{R_s}(>\omega_s)$ was well approximated by
\begin{eqnarray}
P_{R_s}(>\omega_s)\simeq 
\max_{R_L}\left[
\int\d^D\lambda_i\ n_{R_L}(\lambda_i)\ 
V_{\rm caus.}(R_L,R_s,\{\lambda_i\},\omega_s)\right] \, . \label{e:pr1}
\end{eqnarray}
We will now show that the maximum is indeed given by caustics of size of the
order  of $R_s$ and approximate this integral in 2 and 3D.    To  simplify  further \Eq{pr1},   note first  that   the
distribution function of the eigenvalues is peaked in a given geometry (i.e.
$a=1$,   and $  b \simeq 0$   in  3D) for   rare caustics  (large  values of
$\lm$). Therefore the integral in \Eq{pr1} will  be dominated by caustics of
this geometry and the factor $V_{\rm caus.}$ can be taken at this point
while carrying the integration over the other two eigenvalues.  As
a result we have
\begin{eqnarray}
P_{R_s}(>\omega_s)\simeq 
\max_{R_L}\left[
\int_1^{\infty}
\d\lm\ p_{\rm max}(\lm)\ \left({\lm\over\sigma(R_L)}\right)^D\,
{n_0(\lm)\ V_{\rm caus.}(R_L,R_s,\lm,\omega_s)\over
R_L^D}
\right].\label{e:pr2}
\end{eqnarray}
This integral runs  from 1 to  infinity since  the caustics exist  only when
$\lm$ is greater than 1.  The  evaluation of \Eq{pr2} requires insights into
the function    $V_{\rm caus.}$.  Although there  are    no real qualitative
changes between the the 2D and 3D cases, we now proceed with the computation
of \Eq{pr2} by distinguishing the two geometries for the sake of clarity.

\subsection{The 2D statistics}
Recall that the integral \Eq{pr1} will  be dominated by  the rare even tail,
and thus by the lowest value of  $\lm$ that contributes  to the integral. In
other   words,  when  considering  a  given   caustic  characterized  by its
Lagrangian scale $R_{L}$,  one should wait long  enough so that it has grown
sufficiently in  order to contribute after sampling  a vorticity larger than
$\omega_{s}$.  For each  $R_{L }$  therefore corresponds $\lm^{(0)}(R_{L})$,
the lowest value of $\lm$ for which $V_{\rm caus.}$ is non zero:
\begin{eqnarray}
P_{R_s}(>\omega_s)\simeq 
\max_{R_L}\left[
\int_{\lm^{(0)}}^{\infty}
\d\lm\ p_{\rm max}(\lm)\ \left({\lm\over\sigma(R_L)}\right)^2\,
{n_0(\lm)\,V_{\rm caus.}(R_L,R_s,\lm,\omega_s)\over
R_L^2}
\right].\label{e:pr3}
\end{eqnarray}
The lower  bound   $\lm^{(0)}(R_{L})$  is reached as  soon   as $\omega_{\rm
 quad.}$ is larger than $ \pi\  R_s^2\ \omega_s$: the largest possible value
 of the integrated vorticity in  a cell of a given  radius.  It is therefore
 implicitly defined by
\begin{equation}
\disp{\omega_s= { \omega_{\rm quad.} \over \pi\ R_s^2 } \equiv
 \omega_M = f(\Omega)\ 
{R_L^2\over \pi R_s^2}\omega_0\ (\lm^{(0)}(\omega_s,R_{L})-1)^{\alpha}} \, .
\label{e:oms} \end{equation} 
Assuming that $V_{\rm  caus.}$ does not  contain any exponential cutoff, and
assuming  that  $\lm$ is   in   the   rare  event  tail,  \Eq{pr2}  can   be
approximatively re-expressed as
\begin{eqnarray}
&&P_{R_s}(>\omega_s)\simeq 
\max_{R_L}\left[0.56\ \left({\lm^{(0)}\over\sigma(R_L)}\right)^{2}\ 
\exp\left[-{4\over3}\left({\lm^{(0)}\over\sigma(R_L)}\right)^2\right]
{n_0\,V_{\rm caus.}(R_s,R_L,\lm^{(0)},\omega_s)\over R_L^2}
\right],\label{e:pr2b}
\end{eqnarray}
when using \Eq{p2Dlm} for the distribution function of $\lm$, integrating by
part     and    dropping  the     residual  integral    for   large   enough
${\lm^{(0)}/\sigma(R_L)}$ (see Appendix~\ref{s:rare}    for details).   This
maximum with  respect to $R_{L}$ is then  approximated by the minimum of the
argument   of  the  exponential, ${\lm^{{(0)}}(R_L)/\sigma(R_L)}$, where  the
minimum in  the facto taken  with respect to $\lm^{(0)}$ since $\sigma(R_L)$
can be thought of a function of  $\lm^{{(0)}}$ via \Eqs{sigRL} and \Ep{oms}.
This minimum can de facto be expressed independently of $R_s$.  It reads
\begin{equation}   
\lm^{(0)}=\disp{4\over 4-\alpha(n+2)}. \label{e:lmcrit}
\end{equation}
Once $\lm^{(0)}$  is fixed  the geometry of  the
caustic which will contribute most  to $P_{R_s}(>\omega_s)$  is entirely
specified.  The condition
for the existence of a minimum defining $\lm^{(0)}$ is that $\alpha(n+2)<4$,
and it is satisfied for all considered cases (see \Tab{tab1}).  This implies
that  we are  investigating a  r\'egime where  the integral  \Eq{pr2} is not
dominated by arbitrarily rare caustics -- which would have been catastrophic
given the assumptions (note that when $n$ is  too large $\lm^{(0)}$ tend to
be quite large  thus challenging the   validity of quantitative  results
based upon the Zel'dovich approximation).  The resulting value of $R_L$ is
\begin{equation}
R_L 
=\disp{R_s\ \sqrt{{\pi \omega_s \over \omega_0\,f(\Omega)}\,}
\,\left({  4-\alpha\,(n+2) \over \alpha\,(n+2)}\right)^{\alpha/2}}
=f_s\ R_s\ \left({\omega_s\over f(\Omega)}\right)^{1/2}. \label{e:RL2D}
\end{equation}
The  scale factor $f_s$  is given  in \Tab{tab3}  for an  Einstein-de Sitter
universe  ($f(\Omega)=1$)  and different values    of  $n$.  Completing  the
calculation of $P_{R_s}(>\omega_s)$ involves relating the  shape and size of
the caustic for  the adopted value of $\lm^{(0)}$.   These values are
derived
 from the fits    (\Eq{scaling})   and are  given  in  \Tab{tab3}.
\Fig{fig-10} gives $V_{\rm  caus.}$, in units of  the square of $R_L$,  as a
function of the smoothing radius $R_s$.  From \Fig{fig-10} it is easy to see
that
\begin{equation}
     V_{\rm caus.}\simeq V_0\ R_s\,R_L \, , 
\label{e:vc2D}
\end{equation}
for any values and $n$; the corresponding values of $V_0$ are given in Table
 \ref{t:tab3}.    Putting  \Eq{vc2D}  into   \Eq{pr3},   using \Eqs{lmcrit},
 \Ep{RL2D} yields for the sought distribution
\begin{equation}
P_{R_s}(>\omega_s)\simeq 
0.56\,n_0\,V_0\,\left({\lm^{(0)}\over \sigma(R_s)}\right)^2\,
f_s^{n+1}\,\omega_s^{(n+1)/2}\,
\exp\left[-{4\over 3}\,\left({\lm^{(0)}\over\sigma(R_s)}\right)^2\,
f_s^{n+2}\,\omega_s^{(n+2)/2}\right],\label{e:p2Dfinal}
\end{equation}
Note  that the power of $\omega_s$  in the exponential  is rather weak.
The cut-off is nonetheless  strong in the r\'egime of interest because
of the  leading   coefficient.   Equation \Ep{p2Dfinal} is   illustrated  on
\Fig{fig-f17} and discussed in the main text.

\begin{table}
\caption{Parameters of interest for the 2D caustics:
the power index, $n$, the critical time $\lm^{(0)}$,
the  
radial extension $e^{(0)}$, depth $d^{(0)}$ in units of 
$R_L$,scale factor $f_s^{(0)}$  
as well as the values of $n_0$ and
$V_0$ for the critical caustics.}
\label{t:tab3}
\begin{tabular}{@{}lcccccc}
$n$&$\lm^{(0)}$&$d^{(0)}$&$e^{(0)}$&$f_s$&$n_0$&$V_0$\\
-1.5& 1.31 & 0.17 & 1.34 & 0.30 & 0.018 & 0.9 \\
-1& 1.67 & 0.40 & 1.33 & 0.95 & 0.023 & 1.8 \\
-0.5& 2.15 & 0.90 & 1.36 & 1.25 & 0.009 & 3.4
\end{tabular}
\end{table}

\begin{table}
\caption{Parameters of interest for the 3D caustics:
the power index, $n$, the critical times $\lm^\pm$,
the scale factor $f_s^\pm$ in the two r\'egimes
($R_s<e/3$ in parentheses) with 
radial extension $e^{(0)}$, depth $d^{(0)}$  in units of 
$R_L$ 
as well as the values of $n_0$ and
$V_0$ that enter the final expressions.}
\label{t:tab4}
\begin{tabular}{@{}lccccccccc}
$n$
&$\lm^+$&($\lm^-$)&$f_s^+$&($f_s^-$)&$d^{(0)}$&$e^{(0)}$&$n_0$&$V_0$\\
-2.  & 1.41 & (1.47) & 2.46 & (2.09) & 0.18 & 1.04 &0.18 & 0.96 \\
-1.5  & 1.63 & (1.79) & 2.10 & (1.58) & 0.28 & 1.05 &0.14 & 1.84 \\
-1.  & 1.84 & (2.15) & 1.78 & (1.17) & 0.42 & 1.07 &0.064 & 3.18 
\end{tabular}
\end{table}

\begin{figure}
\begin{minipage}[b]{0.45 \textwidth}
\centering \includegraphics[width=3.in]{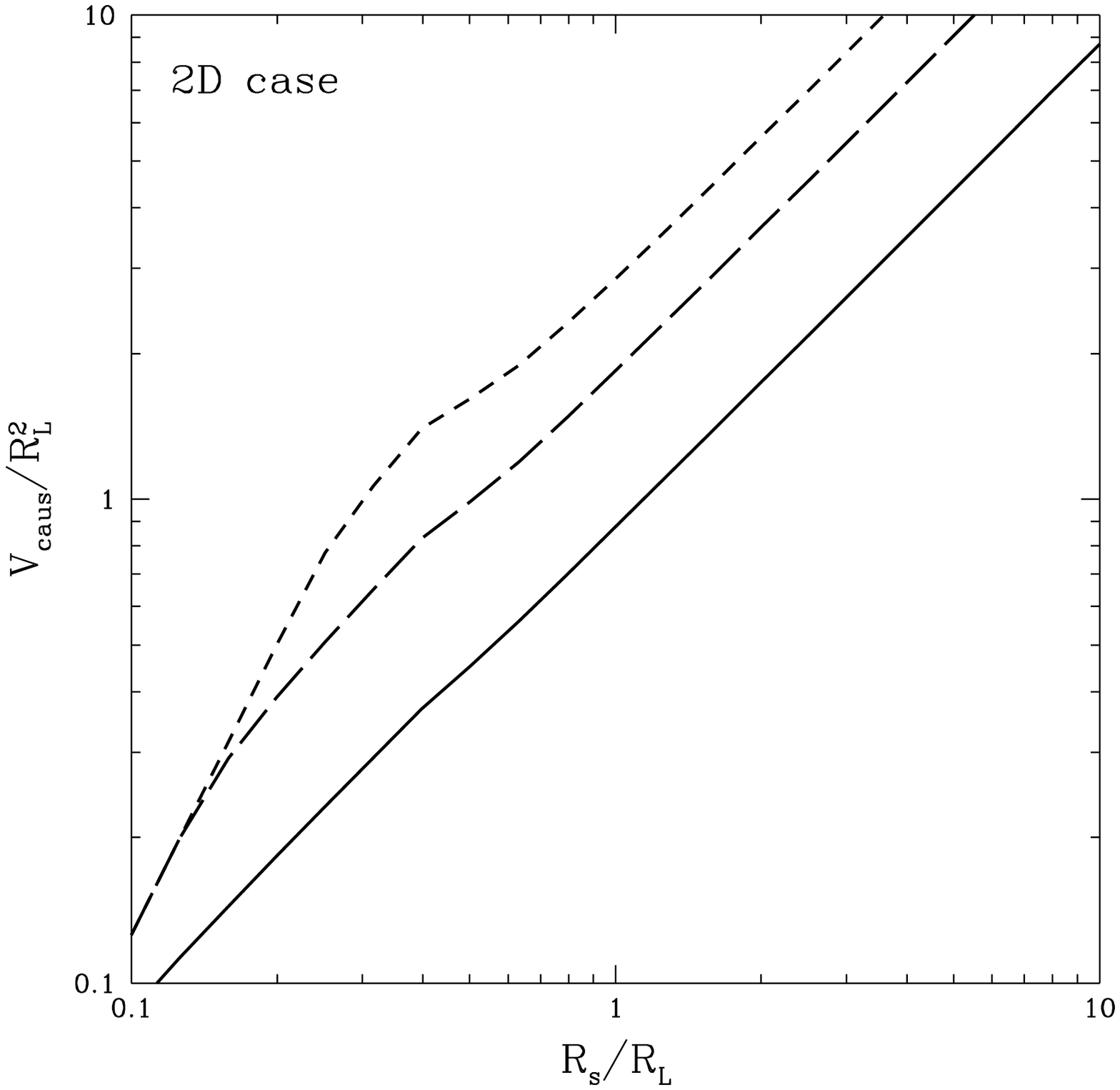}
\caption{The function $V_{\rm caus.}$, in units of the square of $R_L$, as a
function of the  smoothing radius in 2D.  The solid line corresponds  to the
case $n=-1.5$, the  dashed   line to $n=-1$ and   the  long dashed  line  to
$n=-0.5$.  In   all  cases  the  geometry of     the caustic  is   fixed  by
$\lm=\lm^{(0)}$.}
\label{f:fig-11}
\end{minipage}
\begin{minipage}[b]{0.1 \textwidth}
\end{minipage}
\begin{minipage}[b]{0.45 \textwidth}
\centering \includegraphics[width=3.in]{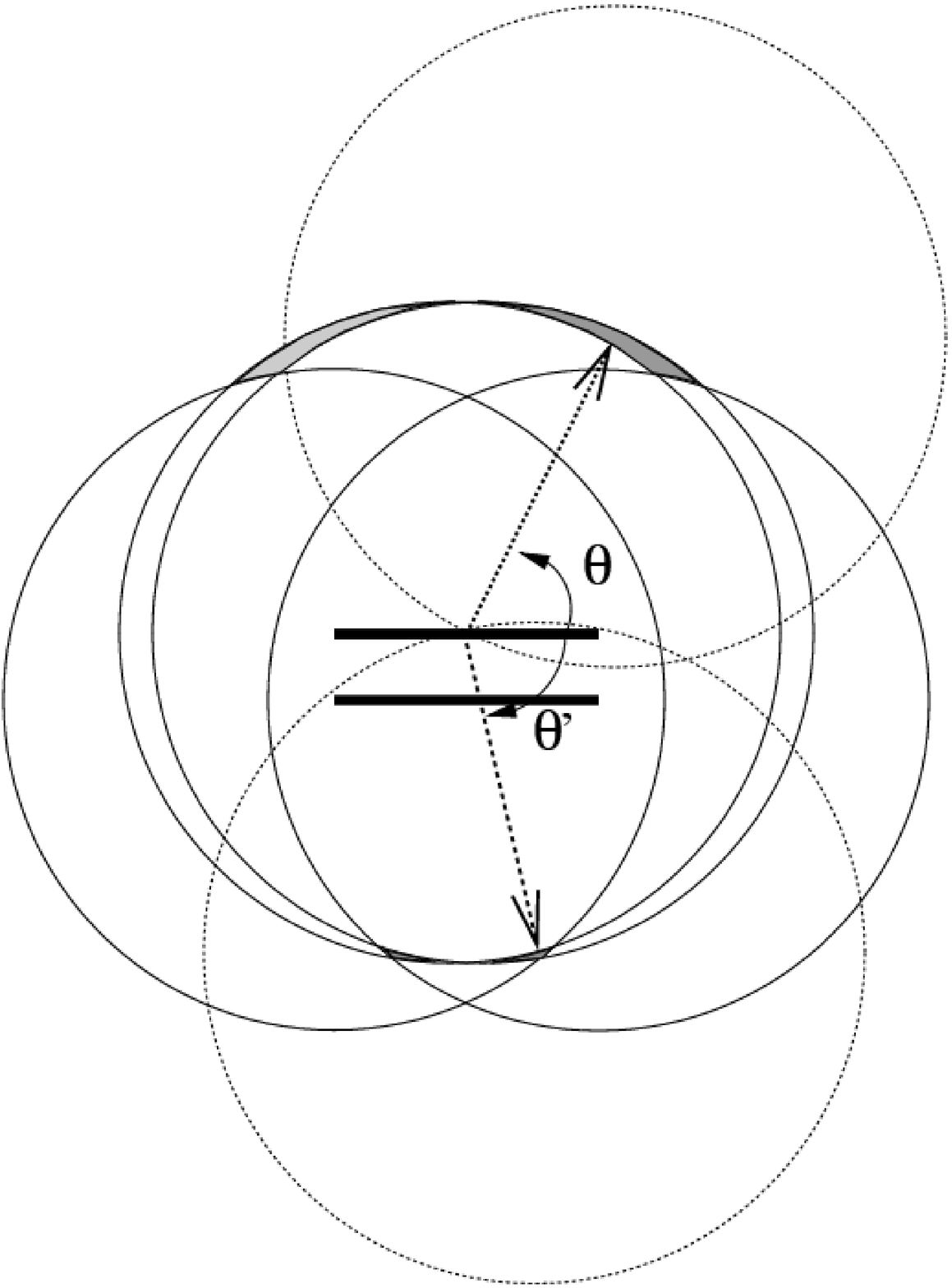}
\caption{The loci  of the centres of  spheres contributing $\omega_s$ in the
 range $[\omega_M^+(1-\epsilon^2/2)$, $\omega_M^+[$. The dashed arrow points
 to  a centre of  such a  sphere, and defines  the running  angle, $\theta$,
 mentioned  in  \Eq{angletheta}.   The two  (cylindrically symmetric) shaded
 regions correspond  to the loci of  the centre of  spheres capturing almost
 half a ring and all or none of the other. Two  examples of such spheres are
 displayed for either case. }
\label{f:figAsymptote}
\end{minipage}
\end{figure}

\subsection{the 3D statistics}

The threshold on  $\lm$, from which  the caustics  start to contribute  at a
given scale    $R_s$  depends on  the adopted    description   for the local
vorticity.   We assume here as  mentioned  in Sect.~\ref{s:rings} that the
total vorticity is localized on  two rings of  radius $e/3$ each, distant of
$2\,d/3$ of  each  other. They  are  assumed to  bear  opposite lineic  (and
uniformly distributed) vorticities; in order to get  a consistent answer for
the integrated vorticity in a quadrant, we should have
\begin{equation}
\omega_{\rm lin.}={3 \omega_{\rm quad.}\over e}.
\end{equation}
The maximum vorticity  that can be  encompassed in a  sphere then depends on
its radius $R_s$.  If $R_s$ is larger than the radius of the rings $e/3$, it
is possible  to have half of a  ring in a sphere (while  the other ring does
not intersect it at all), so that the values of $\lm$  (for which the
maximum vorticity is sampled)
is given by
\begin{equation}
\omega_s={2\,e\,\omega_{lin.}\over3}\,{1\over
{4\pi\over3}R_s^3}\equiv \omega_M^{+}={3\over 2\pi}\omega_0\,(\lm-1)^{\alpha}
f(\Omega)\,{R_L^3\over R_s^3}, \quad{\rm if}\quad R_s>e/3.
\label{e:oms3D1}
\end{equation}
If on the other hand $R_s$ is smaller than $e/3$ then only a fraction of the
half ring can be put in the sphere and we  have instead
\begin{equation}
\omega_s={2\,R_s\,\omega_{lin.}}\,{1\over
{4\pi\over3}R_s^3}\equiv \omega_M^{-}={9\over 2\pi}\,{\omega_0\over e_0}
\,(\lm-1)^{\alpha-\alpha_e}
f(\Omega)\,{R_L^2\over R_s^2}, \quad{\rm if}\quad R_s<e/3.
\label{e:oms3D2}
\end{equation}
Now the local behaviour  of $V_{\rm caus.}$ near its  takeoff value is  well
 represented (as  argued below and demonstrated in Appendix~\ref{s:asymptote}
 for large enough $R_{s}$) as a function of $\lm$ by 
\begin{equation}
V_{\rm caus.}(R_L,R_s,\lm,\omega_s) =
\int \Theta\left[
\omega_{\bf c }\left({\bf c },R_L,R_s,\lm \right)- 
\omega_s \right] 
\, \d^3 c \simeq R_L R_s^2
V_0 (\lm -\lm^{(0)})^{\gamma}  \, , 
\end{equation}
Using   \Eq{p3Dlm} and \Ep{nRl2} for   the  distribution function $\  p_{\rm
max}(\lm)$, changing integration  variable from $u=\lm/\sigma$ to $\lm^{(0)}
+  u /   \lm^{(0)}$ and  dropping  the residual  integral for   large enough
${\lm^{(0)}/\sigma(R_L)}$ (see Appendix~\ref{s:rare} for details) yields for
\Eq{pr3} :
\begin{eqnarray}
&&P_{R_s}(>\omega_s)\simeq 
\max_{R_L}\left[ {6\,n_0\,V_0 \Gamma(\gamma+1) \over 5^{\gamma +1} } \left(
{\lm^{(0)}
 \over\sigma(R_L)}\right)^{4- \gamma}\ 
\exp\left[-{5\over 2}\left({\lm^{(0)}\over\sigma(R_L)}\right)^2
\right] {R_s^2 \over R_L^2} \right]
,\label{e:pr3b}
\end{eqnarray} 
From \Eq{sigRL} and \Ep{oms3D1}, \Ep{oms3D2}, the minimum of the argument of
 the exponential corresponds to:
\begin{equation}
{\lm^{+} \equiv {6 \over 6 -  \alpha (n+3) } 
\quad {\rm if } \quad R_{s}>  e/3 }, 
\quad {\rm and } \quad
{\lm^{-}\equiv {4 \over 4 -  (\alpha-\alpha_e)  (n+3) } 
\quad {\rm if } \quad R_{s} <  e/3 }, \label{e:lcrit3D}
\end{equation}
which assumes that $\alpha(n+3)<6$ (resp. $(\alpha-\alpha_e)(n+3)<4$ ), both
conditions being satisfied for all values  of $n$ considered.  The corresponding
 scaling relations between $R_L$ and $R_s$ are given by
\begin{equation}
R_L     =   f_s^+\ R_s\ \left({\omega_s\over f(\Omega)}\right)^{1/3} 
\quad {\rm if }  \quad { R_{s} > e/3} \, ,   \quad {\rm or} \quad    
R_L     =   f_s^-\ R_s\ \left({\omega_s\over f(\Omega)}\right)^{1/2} 
\quad {\rm if }  \quad { R_{s} < e/3} \, .    
\end{equation}
The scale factors  $f_s^\pm$ -- derived  from the fits (\Eq{scaling}) -- are
given  in \Tab{tab4} for an Einstein-de  Sitter universe ($f(\Omega)=1$) and
different values of  $n$. Interestingly, as long  as  $\omega_s$ is not  too
large the  condition $R_L>e/3$ is always  satisfied.   In practice at scales
of about $10$ to $15h^{-1}$Mpc the measured vorticity $\omega_s$ is expected to be indeed at most of a
few tenth (Bernardeau \& van de Weygaert, 1995). It is therefore always fair
to assume that we are in the r\'egime where $R_s>e/3$ which is the r\'egime
investigated hereafter.

Completing the  calculation of $P_{R_s}(>\omega_s)$  requires evaluating the
corresponding   $n_0$, $\gamma$ and $V_0$.   The  value of $n_0$ is entirely
determined   by  the  geometry   of  the  caustics  and   is   given in  the
Tables~\ref{t:tab1} and~\ref{t:tab2}.   The behaviour  of $V_{\rm caus.}$ as
it departs from zero as a function of  $\omega_s$ for the critical ratios of
$R_s$, $e$ and $d$ is locally well  fitted as a function  of $\omega_s$ by a
power law of the form
\begin{equation}
V_{\rm caus.}(\lm,\omega_s)
\simeq  
U_0\ R_L \, R_s^2
 \left(1- {\omega_s \over \omega_M^+(\lm) } \right)^{\gamma} 
\ \, . \label{e:approxVc}
\end{equation}
where     $\omega_{M}^+$   is   the      threshold   value  of    $\omega_s$
(\Eq{oms3D1}). This expression is valid when $\omega_s$ is close to its threshold
value. On  the critical line,   $\omega_s=\omega_{M}^+$, it is  possible  to
relate the variation of $\lm$ to the  variations of $\omega_s$.  We can then
rewrite \Eq{approxVc} as a function of  the difference between $\lm$ and the
critical value $\lm^{(0)}$, assuming this departure is small,
\begin{equation}
V_{\rm caus.}(\lm,\omega_s)
\simeq   R_L \, R_s^2 V_0\,(\lm -\lm^{(0)})^\gamma
\ \, , 
\quad {\rm with } \quad
V_0 = { U_0\, \alpha^\gamma \over  (\lm^{(0)}-1)^{\gamma}}
\ \, . \label{e:approxVc3}
\end{equation}
Since $R_L/R_s$ is only  a function of $n$ and  $\omega_s$, so are $V_0$ and
$\gamma$.  In practice  we take the  asymptotic values of $V_0$ and $\gamma$
given in Appendix~\ref{s:asymptote} and  corresponding to the limit $R_s \ll
R_L$.     Putting  \Eq{approxVc3}   into     \Eq{pr3b},   using \Eq{lcrit3D}
--\Ep{approxVc}    and  \Ep{vcaustinfinity}   yields    for   the  vorticity
distribution
\begin{equation}
P_{R_s}(>\omega_s) = 0.48 n_0 V_0 
\left({\lm^{(0)}\over \sigma(R_s)}\right)^{7/2}\,
f_s^{(13+7n)\over 4}\,\omega_s^{(13 +7 n) \over 12}\,
\exp\left[-{5\over 2}\,\left({\lm^{(0)}\over\sigma(R_s)}\right)^2\,
f_s^{n+3}\,\omega_s^{(n+3)/3}\right] \, .\label{e:p3Dfinal}
\end{equation}
Equation \Ep{p3Dfinal} is illustrated on \Fig{fig-f18} and discussed in the 
main text.

\section{ Asymptotic behaviour of $V_{\rm Caust.}$ in 3D } \label{s:asymptote}

For large enough $R_s$ we derive here an  asymptotic analytic expression for
$V_{\rm Caust.}$.  Let  us first estimate geometrically  the volume in space
contributing  almost  $\omega_M^+$ to  $V_{\rm Caust.}$.   The corresponding
contribution   is the  sum of  two   volumes given  by  the  shaded area  in
\Fig{figAsymptote},  corresponding to the  loci  of  the centers of  spheres
which  capture  almost half a  ring   and not the    other, or which capture
completely one ring and almost half of  the other.  In the asymptotic limit,
as $e/R_s \to 0$, the element of volume is an infinitely thin strip and both
contributions    become equal    since  $\theta  \to   -\theta'$.   The area
corresponding to these loci can be evaluated algebraically as follow: let us
call $\epsilon$ the  projected ring segment by   which a sampling  sphere of
radius $R_{s}$ fails to encompass a  ring diameter $2  e/3$; it follows that
the ratio of $\omega_s$ to $\omega_M^+$, is given by
\begin{equation}
{\omega_s \over \omega_M^+} = (1-{\epsilon^2 \over 2}) \,  . 
\label{e:epsilon}
\end{equation}
On the  other hand, for  a given  direction  for the  sphere centre given by
$\cos(\theta) \equiv \mu$, within the solid angle $ 2 \pi \d \mu$, the volume
element (encompassed by the two shifted spheres  capturing $\omega_s$ in the
range $[\omega_M^+(1-\epsilon^2/2)$, $\omega_M^+[$) is given by
\begin{equation}
\left( {2 e \over 3} \right) 4 \pi \, R_s^2  \epsilon \sin^2 \theta \d
 \theta=  8 \pi {e\over 3} R_s^2 \epsilon { \sqrt{1-\mu^2}} \, \d \mu \, . 
 \label{e:angletheta}
\end{equation}
Summing over all  possible  directions ({\it  i.e.} before intersecting  the
second ring) yields
\begin{equation}
 8 \pi {e \over 3} R_s^2 \epsilon \int \limits_{\mu_0}^1 { \sqrt{1-\mu^2}} \, \d \mu \,
 \equiv 
 \, 8 \,  \pi \, {e^{(0)} \over 3}  R_L
 \,  R_s^2  \, \epsilon {\cal J} , \quad  {\rm where} \quad
\mu_0 =  \left[1+ { 4 {d^{(0)}}^2 \over  {e^{(0)}}^2 } \right]^{-1/2} \, .
\end{equation}
Accounting for the   summation  over the  two  configurations  (half a  ring
captured or a full + one half ring captured), using \Eq{epsilon} to
eliminate $\epsilon$, we finally
get for large enough $R_{s}$
\begin{equation}
V_{\rm Caust.} = 16 \sqrt{2} \pi R_L \, R_s^2  \,{ e^{(0)} \over 3} \,
\left(1- {\omega_s \over \omega_M^+ } \right)^{1/2}    \, , \quad
  {\rm therefore} \quad \gamma_\infty = {1\over 2}  \quad {\rm and} \quad
U_0^\infty = 16 \sqrt{2} \pi   { e^{(0)} \over 3} \, {\cal J} \, .
\label{e:vcaustinfinity}
\end{equation}

\section{ Rare event approximation} \label{s:rare}

Consider an integral of the form 
\begin{equation}
{\cal I}= \int \limits_a^\infty x^\beta (x-a)^\gamma \exp(- b x^2) \, \d x 
\, .
\label{e:calI}
\end{equation}
Changing variable to $x=a + u/(2  a b ) $ \Eq{calI} 
reads
\begin{equation}
{\cal I}= {1 \over 2 a b } \exp(- b a^2)  \int \limits_0^\infty \left( {u \over 2 a b}
 \right)^\gamma a^\beta \left[ \left(1+ 
{u \over 2  a^2 b }\right)^\beta 
\exp\left(-{ u^2 \over 2 b a^2 }\right) \right] \,  \exp(- u)  \, \d u 
\, .
\label{e:calI2}
\end{equation}
For large enough $a$
the square brace in \Eq{calI2} is well approximated by $1$ 
yielding for \Eq{calI2}
\begin{equation}
{\cal I}= {a^{\beta -\gamma -1} \over (2  b)^{\gamma+1}
 }  \Gamma({\gamma +1}) \, \exp(- b a^2) \, .
\label{e:calI3}
\end{equation}
\Eq{pr2b}   is  a  special  case of  \Eq{calI}   with  $x= \lm/\sigma$, $a  =
\lm^{(0)}$,  $\gamma=0$,   $\beta  =  3$  and    $b  =4/3$, while \Eq{pr3b}
corresponds to $\beta = 5$,  and $b =5/2$. Note that
the $\gamma =0$ approximant can be deduced directly by integration by
parts.

\end{document}